\documentclass[10pt,natbib]{iopart}

\usepackage{iopams}

\expandafter\let\csname equation*\endcsname\relax

\expandafter\let\csname endequation*\endcsname\relax

\usepackage{amsmath}

\usepackage{amsfonts}

\DeclareMathAlphabet{\mathpzc}{OT1}{pzc}{m}{it}
\newcommand{\Tt}[1]{\boldsymbol{#1}}
\newcommand{\Ttt}[1]{\boldsymbol{#1}}

\newcommand{\sLNsq}{\mathpzc{L}}
\newcommand{\mc}[1]{\mathcal{#1}}
\renewcommand{\ms}[1]{\mathscr{#1}}

\newcommand{\bdX}{\bullet}
\newcommand{\bdR}{\circ}
\newcommand{\sqkX}{\mathbin{\substack{\bdR}{\hspace{-.2ex}}\substack{\bdX\\ \bdX}}}
\newcommand{\sqkR}{\mathbin{\substack{\bdR}{\hspace{-.2ex}}\substack{\bdR\\ \bdR}}}
\newcommand{\sqkXR}{\mathbin{\substack{\bdR}{\hspace{-.2ex}}\substack{\bdX\\ \bdR}}}
\newcommand{\sqcX}{\mathbin{\substack{\bdX}{\hspace{-.2ex}}\substack{\bdX\\ \bdX}}}
\newcommand{\sqcR}{\mathbin{\substack{\bdX}{\hspace{-.2ex}}\substack{\bdR\\ \bdR}}}
\newcommand{\sqcXR}{\mathbin{\substack{\bdX}{\hspace{-.2ex}}\substack{\bdX\\ \bdR}}}
\newcommand{\velG}{{v}}\newcommand{\umode}{{Z}}\newcommand{\radius}{{R}}

\newcommand{\seigen}{e}\renewcommand{\mathtt}[1]{{#1}}

\newcommand{\paperone}{{Part 
\bf{I}}}

\newcommand{\Z}{\mathbb{Z}}
\newcommand{\C}{\mathbb{C}}
\usepackage{graphicx}

\usepackage{mathtools}
\usepackage{cases}

\usepackage{graphicx} 
\usepackage{dcolumn} 
\usepackage{bm} 

\usepackage[mathscr]{euscript}

\usepackage{mathtools}
\usepackage{environ}

\usepackage{xcolor}
\usepackage{color}

\usepackage{upgreek}
\usepackage{xcolor}
\usepackage{color}

\usepackage{mathtools}
\usepackage{cases}
\usepackage[colorlinks=true,linkcolor=blue, citecolor=brown, urlcolor=violet]{hyperref}

\usepackage{graphicx}
\usepackage{bm}

\usepackage{lineno}
\setpagewiselinenumbers

\makeatletter
\newcommand\RedeclareMathOperator{ \@ifstar{\def\rmo@s{m}\rmo@redeclare}{\def\rmo@s{o}\rmo@redeclare}}
\newcommand\rmo@redeclare[2]{\begingroup \escapechar\m@ne\xdef\@gtempa{{\string#1}}\endgroup
 \expandafter\@ifundefined\@gtempa
 {\@latex@error{\noexpand#1undefined}\@ehc} \relax
 \expandafter\rmo@declmathop\rmo@s{#1}{#2}}
\newcommand\rmo@declmathop[3]{ \DeclareRobustCommand{#2}{\qopname\newmcodes@#1{#3}}}
\@onlypreamble\RedeclareMathOperator
\makeatother

\newcommand{\pd}[2]{\frac{\partial#1}{\partial#2}}
\newcommand*{\bfrac}[2]{\genfrac{}{}{0pt}{}{\raisebox{-.3em}{\scriptsize$#1$}}{\raisebox{.4em}{\scriptsize$#2$}}}

\newcommand{\beqan}{\begin{eqnarray}}
\newcommand{\eeqan}{\end{eqnarray}}
\newcommand{\beqans}{\begin{subequations}\begin{eqnarray}}
\newcommand{\eeqans}[1]{\end{eqnarray}\label{#1}\end{subequations}}

\usepackage{environ}
\NewEnviron{eqn}{\begin{equation}\begin{split}
 \BODY
\end{split}\end{equation}
}

\usepackage[mathscr]{euscript}

\begin{document}

\title[Conductance of discrete bifurcated waveguides]{Conductance of discrete bifurcated waveguides as three terminal junctions}

\author{Basant Lal Sharma}

\address{Department of Mechanical Engineering, Indian Institute of Technology Kanpur, Kanpur, U. P. 208016, India}
\ead{bls@iitk.ac.in}
\vspace{10pt}
\begin{indented}
\item[]\today
\end{indented}

\begin{abstract}
{{An expression for the transmission matrix based conductance is provided for the propagation of scalar waves in certain bifurcated discrete waveguides using the paradigm of a three-terminal Landauer--B{\"{u}}ttiker junction. It is found that the conductance across the terminals of bifurcated branches forming a sharp corner, interpreted as a controller of `leakage' flux, can be tuned by manipulating the number of channels and the type of lateral confinement. Natural applications in engineering and science arise in the context of nanoscale transport involving elastic, phononic, or electronic waves. 
{In particular, the paper includes} a discussion of temperature dependent thermal conductance, assuming only the contribution of out-of-plane phonons, along with some graphical illustrations.}}
\end{abstract}

%
%
%
%
%


\maketitle
\linenumbers

\section{Introduction}
\label{intro}
Many current technological blueprints 
often involve phononic \cite{Cahill1}, photonic \cite{Kosevichmulti,GalliBelotti}, and electronic \cite{Zwierzycki2008,Sorensen2009}, transport in nanostructures. The critical issues are often analyzed using the governing equations which have similar mathematical structure.
In this paper, one such prototype discrete model of a three-terminal junction \cite{Csontos2002} is used to investigate phonon scattering and transmission. 
The structure is based on a classical problem in wave mechanics, in the continuum framework, which is also called the bifurcated parallel waveguide problem \cite{marcuvitz1951waveguide,HurdGruenberg,LuneburgHurdbif}. With Dirichlet boundary conditions on the outer walls and on the inserted half-plane, positioned asymmetrically in general, edge scattering of the incident {wave} gives rise to the propagating and evanescent waveguide modes in all the three regions: {above as well as below the bifurcation and the intact region}. 
Incidence from {narrow} waveguide portions, as well as the replacement of Dirichlet conditions by Neumann boundary conditions follow analogous treatments \cite{mittra1971analytical}. 

\begin{figure}[ht]
\centering
{\includegraphics[width=.8\linewidth]{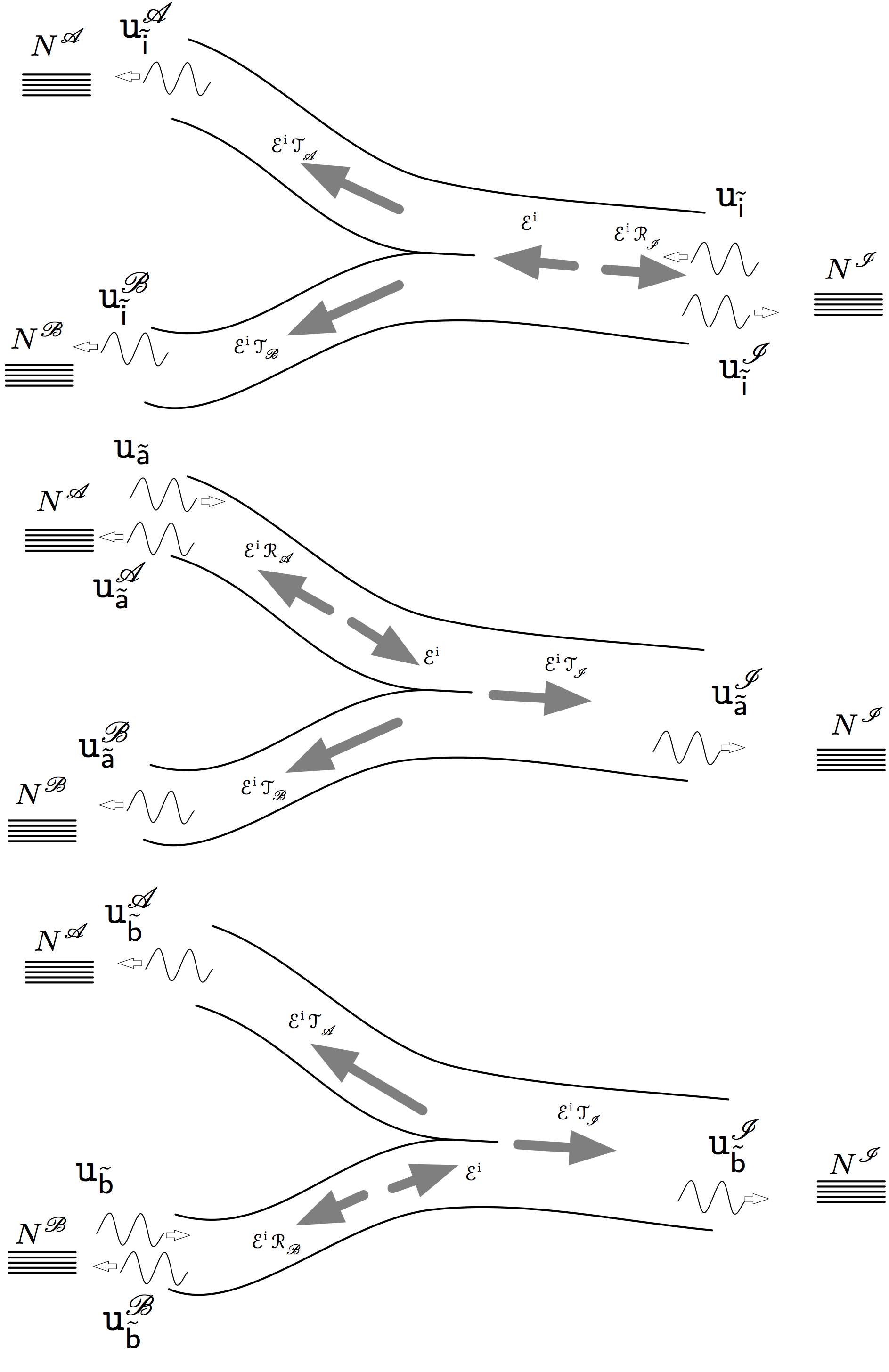}}
\caption{Schematic illustration of the three directions of incidence and associated parameters, {such as, channels in terminals, indices for channels, symbol for asymptotic states, etc}, for the bifurcated waveguides.
{The subscript on ${\mathtt{u}}$ denotes the channel for the incident wave while the superscript denotes the portion into which the wave field is asymptotically evaluated.}}
\label{conductance_bifurcated_sq}
\end{figure}

The prototype model of this paper consists of two-dimensional square lattice waveguides, alternatively interpreted as coupled one-dimensional discrete chains.
The `perfect' waveguides are joined together by a 
`bifurcation'-shaped junction allowing maximum ${\mathtt{N}}$ channels (due to 1D- chains) in the terminal on one side and a set of maximum ${\mathtt{N}_{\mathfrak{a}}}$ and ${\mathtt{N}_{\mathfrak{b}}}$ channels on the two terminals on other side of bifurcation (as schematically shown in Fig. \ref{conductance_bifurcated_sq}). In case of crack induced bifurcation, referred as type I, ${\mathtt{N}}={\mathtt{N}_{\mathfrak{a}}}+{\mathtt{N}_{\mathfrak{b}}}$ while for the rigid constraint induced bifurcation, referred as type II, ${\mathtt{N}}={\mathtt{N}_{\mathfrak{a}}}+{\mathtt{N}_{\mathfrak{b}}}+1$. 
The time harmonic response of such three terminal junction 
exhibits several remarkable properties. 
At a given frequency, 
there are $N^{{\mc{I}}},$ $N^{{\mc{A}}}$, and $N^{{\mc{B}}}$ propagating wave modes (channels) in the three terminals, {while the incident wave mode can occur in any of the three terminals}
(Fig. \ref{conductance_bifurcated_sq}).
The existing theoretical approach using mode-matching \cite{Csontos2002}, {in conjunction} with the {provision} of an exact solution \cite{Bls9s} (henceforth, the same \cite{Bls9s} 
will be referred as \paperone{}) of the system, allows to obtain the scattering matrix for such three terminal junctions.
In fact, it is found that the three terminal junction acts as an effective phonon splitter whose characteristics can be controlled by varying its structural parameters, namely ${\mathtt{N}_{\mathfrak{a}}}$ and ${\mathtt{N}_{\mathfrak{b}}}$. For example, there exist frequency intervals in the pass band of the bulk lattice structure where the reflectance is negligible, as well as those regions where the transmittance is very small, for waves incident from the intact portion. It can be envisaged that the techniques and results of the paper assume qualitative relevance to more complex {physical} systems. \cite{marcuvitz1951waveguide,collin1991field,BurnsMilton,kokubo2011waveguide}.

\begin{figure*}[ht!]
\centering
{\includegraphics[width=\linewidth]{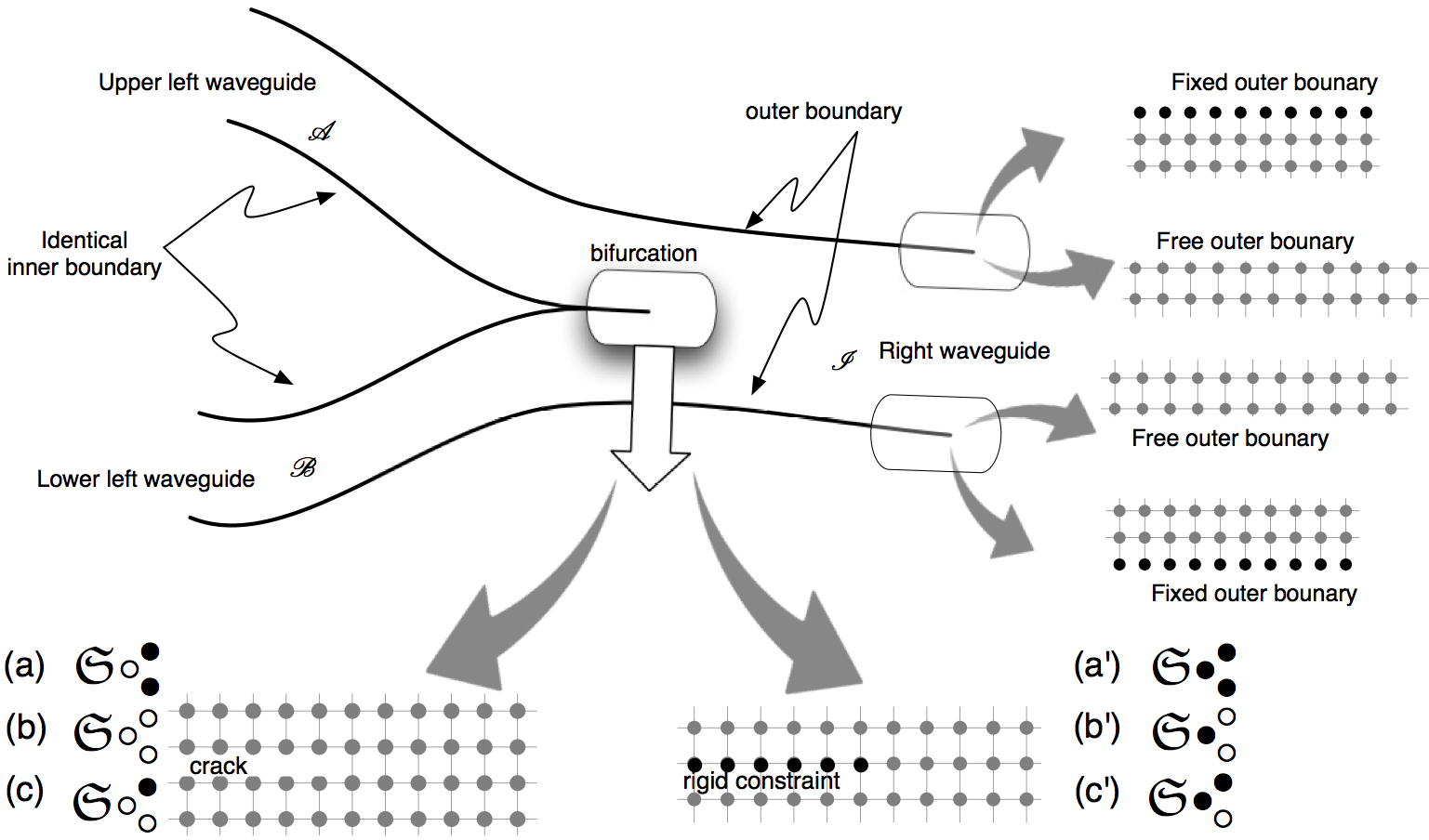}}
\caption{Bifurcated waveguides type I (a, b, c) of square lattice structure (induced by a semi-infinite crack, i.e. `discrete Neumann') 
and type II (a${}^\prime$, b${}^\prime$, c${}^\prime$) (induced by a semi-infinite rigid constraint, i.e. `discrete Dirichlet').}
\label{latticestrip_sq_BCs_asymm_semiinfdefect}
\end{figure*}

\paragraph*{Outline}
The second section captures the {details for} reflectance and transmittance characteristics
{based on the asymptotic expansion of exact solution and mode matching method}. 
The {$3\times 3$} conductance matrix {comprising the appropriate sums of transmission coefficients, following Landauer--B{\"{u}}ttiker formalism}, is elaborated in the third section.
The fourth section is devoted to some critical observations concerning symmetrically located bifurcation, thermal transport {assuming only the contribution of out-of-plane phonons}, and the electronic transport. 
Concluding remarks appear in the end, {while} additional details are provided in three appendices.

\section{Reflection and transmission coefficients in bifurcated waveguides}
\label{latticemodel}
Let
${{\mathfrak{S}_{{\mathtt{N}}}}}$ denote a 
two-dimensional strip of the square lattice strip with ${\mathtt{N}}$ number of rows {of particles.
An out-of-plane motion of the particles is considered assuming each particle has unit mass.} An interaction with atmost four nearest neighbours, through linearly elastic identical (massless) bonds with a spring constant $1/{\mathrm{b}}^2$, {leads to a quasi-one dimensional physical system of particles that allows propagation of waves.}
{Let} the (complex valued) out-of-plane displacement, of a {typical} particle with {\em lattice coordinates} $({\mathtt{x}}, {\mathtt{y}})\in{{\Z^2}}$, be denoted by ${\mathtt{u}}_{{\mathtt{x}}, {\mathtt{y}}}$. 
The equation of motion of 
away from 
{boundaries is easily found to be}
${\mathrm{b}}^{2}\ddot{\mathtt{u}}_{{\mathtt{x}}, {\mathtt{y}}}=\triangle{\mathtt{u}}_{{\mathtt{x}}, {\mathtt{y}}},$
where $\triangle{\mathtt{u}}_{{\mathtt{x}}, {\mathtt{y}}}{:=}{\mathtt{u}}_{{\mathtt{x}}+1, {\mathtt{y}}}+{\mathtt{u}}_{{\mathtt{x}}-1, {\mathtt{y}}}+{\mathtt{u}}_{{\mathtt{x}}, {\mathtt{y}}+1}+{\mathtt{u}}_{{\mathtt{x}}, {\mathtt{y}}-1}-4{\mathtt{u}}_{{\mathtt{x}}, {\mathtt{y}}}.$
{On the other hand,} at the upper and lower outer boundary either $\text{(1) }{\mathrm{b}}^{2}\ddot{\mathtt{u}}_{{{\mathtt{x}}}, {{\mathtt{y}}}}
={\mathtt{u}}_{{{\mathtt{x}}}+1, {{\mathtt{y}}}}+{\mathtt{u}}_{{{\mathtt{x}}}-1, {{\mathtt{y}}}}+{\mathtt{u}}_{{{\mathtt{x}}}, {{\mathtt{y}}}-1}-4{\mathtt{u}}_{{{\mathtt{x}}}, {{\mathtt{y}}}}
\text{ or (2) }{\mathrm{b}}^{2}\ddot{\mathtt{u}}_{{{\mathtt{x}}}, {{\mathtt{y}}}}={\mathtt{u}}_{{{\mathtt{x}}}+1, {{\mathtt{y}}}}+{\mathtt{u}}_{{{\mathtt{x}}}-1, {{\mathtt{y}}}}+{\mathtt{u}}_{{{\mathtt{x}}}, {{\mathtt{y}}}-1}-3{\mathtt{u}}_{{{\mathtt{x}}}, {{\mathtt{y}}}}
\text{ and either (3) }{\mathrm{b}}^{2}\ddot{\mathtt{u}}_{{{\mathtt{x}}}, {{\mathtt{y}}}}
={\mathtt{u}}_{{{\mathtt{x}}}+1, {{\mathtt{y}}}}+{\mathtt{u}}_{{{\mathtt{x}}}-1, {{\mathtt{y}}}}+{\mathtt{u}}_{{{\mathtt{x}}}, {{\mathtt{y}}}+1}-4{\mathtt{u}}_{{{\mathtt{x}}}, {{\mathtt{y}}}}
\text{ or (4) }{\mathrm{b}}^{2}\ddot{\mathtt{u}}_{{{\mathtt{x}}}, {{\mathtt{y}}}}={\mathtt{u}}_{{{\mathtt{x}}}+1, {{\mathtt{y}}}}+{\mathtt{u}}_{{{\mathtt{x}}}-1, {{\mathtt{y}}}}+{\mathtt{u}}_{{{\mathtt{x}}}, {{\mathtt{y}}}+1}-3{\mathtt{u}}_{{{\mathtt{x}}}, {{\mathtt{y}}}}$
respectively, {corresponding to the free boundary condition or that arising due to attachment to a fixed wall.} 
{In case of type I bifurcation, crack corresponds to broken bonds between $\mathtt{y}=0$ and $\mathtt{y}=-1$, while type II bifurcation involves fixed sites at $\mathtt{y}=0$.}
{It is clear that} on the crack faces (resp. on rows adjacent to constraint) (2) holds at ${\mathtt{y}}=-1$ (resp. (1) at ${\mathtt{y}}=-1$) and (4) at ${\mathtt{y}}=0$ (resp. (3) at ${\mathtt{y}}=1$) for ${\mathtt{x}}<0$. A schematic illustration {of the different kinds of boundaries} is provided in Fig. \ref{latticestrip_sq_BCs_asymm_semiinfdefect}.

Suppose ${\mathtt{u}}^{i}$ {denotes} {\em incident wave mode} with frequency $\omega$ and a {\em lattice wave number} ${\upkappa}_x$ along ${\mathtt{x}}$. 
Due to the {confined geometry, ${\mathtt{u}}^{i}$ (any of the waves ${\mathtt{u}}_{{\tilde{{{\mathsf{i}}}}}}$, or ${\mathtt{u}}_{{\tilde{{\mathsf{b}}}}_{\mathfrak{a}}}$, or ${\mathtt{u}}_{{\tilde{{{\mathsf{i}}}}}_{\mathfrak{b}}}$, {following the notation for channel indices} as schematically shown in Fig. \ref{conductance_bifurcated_sq}) {is naturally expressed in terms of a normal mode \cite{Bls9} ${\seigen}_{({\kappa}^{i})}\in\C^{{\mathtt{N}}}$ (indexed by ${\kappa}^{i}$), i.e.}
\footnote{\label{note_npp}For convenience of notation during the application of the normal mode analysis \cite{Bls9}, etc, ${\nu}$ is used in place of ${{\mathtt{y}}}$ such that the lattice row at ${{\mathtt{y}}}$ {is} mapped to ${\nu}={{\mathtt{y}}}+{{\mathtt{N}_{\mathfrak{b}}}}+1,$ for example, lowest row at ${{\mathtt{y}}}=-{{\mathtt{N}_{\mathfrak{b}}}}$ is mapped to ${\nu}=1$.}
\begin{eqn}
{\mathtt{u}}_{{\mathtt{x}}, {\mathtt{y}}}^{i}{:=}{{\mathrm{A}}}{\seigen}_{({\kappa}^{i}){{\nu}}}e^{-i{\upkappa}_x {\mathtt{x}}-i\omega t}, ({\mathtt{x}}, {\mathtt{y}})\in{{\Z^2}N}, 
\label{uinc_sq}
\end{eqn}
where ${{\mathrm{A}}}\in\C$ is constant.} 
Let ${\upomega}$ be defined as ${\upomega}{:=} {\mathrm{b}}\omega.$ Indeed, 
${\upomega}$ and ${\upkappa}_x$ satisfy the well known dispersion relation for square lattice, i.e.
${\upomega}^2=4(\sin^2{\frac{1}{2}}{\upkappa}_x+\sin^2{\frac{1}{2}}{{\upeta}_{{\kappa}^{i}}}), {\upkappa}_x\in [-\pi, \pi],$
where the expression for ${{\upeta}_{{\kappa}^{i}}}$ {depends} on outer boundary conditions \cite{Bls9}. 
In the remaining text, {the factor} $e^{-i\omega t}$ is suppressed.
The total displacement field ${\mathtt{u}}^{{t}}$ ($={\mathtt{u}}_{{\mathtt{x}}, {\mathtt{y}}}^{i}+{\mathtt{u}}_{{\mathtt{x}}, {\mathtt{y}}}$) satisfies the discrete Helmholtz equation
\begin{eqn}
\triangle{\mathtt{u}}^{{t}}_{{\mathtt{x}}, {\mathtt{y}}}+{\upomega}^2{\mathtt{u}}^{{t}}_{{\mathtt{x}}, {\mathtt{y}}}=0, \\
\text{ with }
{\upomega}={\upomega}_1+i{\upomega}_2, {\upomega}_2>0.
\label{dHelmholtz_sq}
\end{eqn}
away from any outer boundary of the waveguides or the bifurcation (Fig. \ref{latticestrip_sq_BCs_asymm_semiinfdefect}). 
The complete solution for all three types of incidences, as schematically shown in Fig. \ref{conductance_bifurcated_sq}, can be obtained 
{using the Wiener-Hopf method}. {In particular,} \paperone{}{} provides extensive analysis for the wave incident from the cracked portion of the waveguide, i.e. the top of Fig. \ref{conductance_bifurcated_sq}, while a brief discussion {also appears in its} \S7.4 for the incidence from the bifurcated portion, i.e. terminals ${\mc{A}}, {\mc{B}}$. See {Appendix \ref{app_exactsol} for the detailed analysis of the latter.}

{Using the exact solution, it is found that}
the total (far-field) displacement field for the type I bifurcation with incidence from the right side ({same as Eq. (4.6) of \paperone{}}), i.e. terminal ${\mc{I}}$, is given by
\begin{eqn}
{\mathtt{u}}^{{t}}_{{\mathtt{x}}, {\mathtt{y}}}&\sim{{\mathrm{A}}}{\seigen}_{({{\kappa}^{i}}){{\nu}}}{{z}}_{{P}}^{{\mathtt{x}}}+{{\mathrm{A}}}\frac{{\ms{D}}_+({{z}}_{{P}})}{{\ms{N}}_+({{z}}_{{P}})}\sum\nolimits_{{{z}}\in{{\umode}}^+_{{\mc{I}}}}
\\&
\frac{({\seigen}_{({{\kappa}^{i}}){{{\mathtt{N}_{\mathfrak{b}}}+1}}}-{\seigen}_{({{\kappa}^{i}}){{{\mathtt{N}_{\mathfrak{b}}}}}}){\seigen}_{({\kappa}_{{z}}){{\nu}}}{{z}}^{{\mathtt{x}}}}{{\seigen}_{({\kappa}_{{z}}){{\mathtt{N}_{\mathfrak{b}}}}+1}-{\seigen}_{({\kappa}_{{z}}){{\mathtt{N}_{\mathfrak{b}}}}}}\frac{1}{{{z}}-{{z}}_{{P}}} \frac{{\ms{N}}_+({{z}})}{{\ms{D}}'_+({{z}})}\\
{\mathtt{u}}^{{t}}_{{\mathtt{x}}, {\mathtt{y}}}&\sim{{\mathrm{A}}}\frac{{\ms{D}}_+({{z}}_{{P}})}{{\ms{N}}_+({{z}}_{{P}})}\sum\nolimits_{{{z}}\in{{\umode}}^-_{{\mc{A}}{\mc{B}}}}\frac{({\seigen}_{({{\kappa}^{i}}){{{\mathtt{N}_{\mathfrak{b}}}+1}}}-{\seigen}_{({{\kappa}^{i}}){{{\mathtt{N}_{\mathfrak{b}}}}}}){\seigen}_{({\kappa}_{{z}}){{\nu}}}{{z}}^{{\mathtt{x}}}}{{\seigen}_{({\kappa}_{{z}}){{\mathtt{N}_{\mathfrak{b}}}}+1}-{\seigen}_{({\kappa}_{{z}}){{\mathtt{N}_{\mathfrak{b}}}}}}
\\&
\frac{1}{{{z}}-{{z}}_{{P}}} \frac{{\ms{D}}_-({{z}})}{{\ms{N}}'_-({{z}})},
\label{farfield_k_sq}
\end{eqn}
as ${\mathtt{x}}\to+\infty$ and ${\mathtt{x}}\to-\infty$, respectively, {where ${\seigen}_{({\kappa}_{{z}})}$ refer to the outgoing modes and other entities such as ${{\umode}}_{\cdot}^\pm$, ${\ms{N}}$, ${\ms{D}}$ (${\ms{N}}_\pm, {\ms{D}}_\pm$ are the multiplicative Wiener-Hopf factors of ${\ms{N}}$, ${\ms{D}}$, respectively), etc, are stated below.}
The total displacement field for the type II bifurcation with incidence
({same as Eq. (4.9) of \paperone{}}) from the terminal ${\mc{I}}$ is given by
\begin{eqn}
{\mathtt{u}}^{{t}}_{{\mathtt{x}}, {\mathtt{y}}}&\sim{{\mathrm{A}}}{\seigen}_{({{\kappa}^{i}}){{\nu}}}{{z}}_{{P}}^{{\mathtt{x}}}+{{\mathrm{A}}}\frac{{\ms{N}}_+({{z}}_{{P}})}{{\ms{D}}_+({{z}}_{{P}})}\sum\nolimits_{{{z}}\in{{\umode}}^+_{{\mc{I}}}}
\\&
\frac{{{\mathpzc{Q}}}({{z}}_{{P}}){\seigen}_{({{\kappa}^{i}}){{\mathtt{N}_{\mathfrak{b}}}+1}}}{{{\mathpzc{Q}}}({{z}}){\seigen}_{({\kappa}_{{z}}){{\mathtt{N}_{\mathfrak{b}}}+1}}}\frac{1}{{{z}}_{{P}}-{{z}}_{{\mathpzc{q}}}^{-1}}\frac{{{z}}-{{z}}_{{\mathpzc{q}}}^{-1}}{{{z}}-{{z}}_{{P}}}\frac{{\ms{D}}_+({{z}})}{{\ms{N}}'_+({{z}})}{\seigen}_{({\kappa}_{{z}}){{\nu}}}{{z}}^{{\mathtt{x}}},\\{\mathtt{u}}^{{t}}_{{\mathtt{x}}, {\mathtt{y}}}&\sim{{\mathrm{A}}}\frac{{\ms{N}}_+({{z}}_{{P}})}{{\ms{D}}_+({{z}}_{{P}})}\sum\nolimits_{{{z}}\in{{\umode}}^-_{{\mc{A}}{\mc{B}}}}\frac{{{\mathpzc{Q}}}({{z}}_{{P}}){\seigen}_{({{\kappa}^{i}}){{\mathtt{N}_{\mathfrak{b}}}+1}}}{{\seigen}_{({\kappa}_{{z}}){{\mathtt{N}_{\mathfrak{b}}}}+2}+{\seigen}_{({\kappa}_{{z}}){{\mathtt{N}_{\mathfrak{b}}}}}}
\\&
\frac{1}{{{z}}_{{P}}-{{z}}_{{\mathpzc{q}}}^{-1}}\frac{{{z}}-{{z}}_{{\mathpzc{q}}}^{-1}}{{{z}}-{{z}}_{{P}}}\frac{{\ms{N}}_-({{z}})}{{\ms{D}}'_-({{z}})}{\seigen}_{({\kappa}_{{z}}){{\nu}}}{{z}}^{{\mathtt{x}}},
\label{farfield_c_sq}
\end{eqn}
as ${\mathtt{x}}\to+\infty$ and ${\mathtt{x}}\to-\infty$, respectively. Above can be identified with \eqref{assumedfarfield_sq},
assuming ${{z}}_{{P}}$ (${:=} e^{-i{\upkappa}_x}$) corresponds to the index ${\tilde{{{\mathsf{i}}}}}$.
The key elements of the Wiener--Hopf formulation are summarized in Table \ref{bifurcatedstripbc_sq_ND} and \ref{bifurcatedstripbc_sq_ND_c} ({The superscript ${}^{\dagger}$ refers to 
\paperone{}, {and ${\mathpzc{H}}={\mathpzc{Q}}-2=2({\vartheta}-1),$ with ${\mathpzc{Q}}$ defined in \eqref{dHelmholtzF_sq}.} Recall that ${\mathtt{N}}={\mathtt{N}_{\mathfrak{a}}}+{\mathtt{N}_{\mathfrak{b}}}$ for the 
type I bifurcation.
and ${\mathtt{N}}={\mathtt{N}_{\mathfrak{a}}}+{\mathtt{N}_{\mathfrak{b}}}+1$ for the type II bifurcation.}), as the results based on them have used in this paper; in particular, the expressions for ${\ms{N}}$ and ${\ms{D}}$. 

In \eqref{farfield_k_sq}, \eqref{farfield_c_sq}, the sets of ${z}_{{\ast}}$ ($=e^{-i{\upxi}_{{\ast}}}$), describing the propagating waves of the general form \eqref{uinc_sq}, corresponding to outgoing waves are
\begin{eqn}
{{\umode}}_{{\mc{I}}}^+&=\begin{cases}
\{{z}_{{\ast}}\in\mathbb{T} \big| {{{\ms{D}}}_+({{z}}_{{\ast}})=0}\},\text{ type I}\\
\{{z}_{{\ast}}\in\mathbb{T} \big| {{{\ms{N}}}_+({{z}}_{{\ast}})=0}\},\text{ type II}\\
\end{cases},\\
 {{\umode}}^{-}_{{\mc{A}}{\mc{B}}}&={{\umode}}^-_{{\mc{A}}}\cup{{\umode}}^-_{{\mc{B}}}, \\
\text{where }
{{\umode}}^-_{{\mc{A}}}&=\begin{cases}
\{{z}_{{\ast}}\in\mathbb{T} \big| {{{{\ms{N}}}_{{\mathfrak{a}}-}}({{z}}_{{\ast}})=0}\}, \text{ type I}\\
\{{z}_{{\ast}}\in\mathbb{T} \big| {{\mathring{{\ms{D}}}_{{\mathfrak{a}}-}}({{z}}_{{\ast}})=0}\}, \text{ type II}\\
\end{cases},\\
{{\umode}}^-_{{\mc{B}}}&=\begin{cases}
\{{z}_{{\ast}}\in\mathbb{T} \big| {{{{\ms{N}}}_{{\mathfrak{b}}-}}({{z}}_{{\ast}})=0}\}, \text{ type I}\\
\{{z}_{{\ast}}\in\mathbb{T} \big| {{\mathring{{\ms{D}}}_{{\mathfrak{b}}-}}({{z}}_{{\ast}})=0}\}, \text{ type II}\\
\end{cases}.
\label{Zer_sq}
\end{eqn}
Correspondingly, the sets of ${z}_{{P}}$ corresponding to incoming waves are
\begin{eqn}
{{\umode}}_{{\mc{I}}}^-=\overline{{{\umode}}_{{\mc{I}}}^+}, {{\umode}}^{+}_{{\mc{A}}{\mc{B}}}={{\umode}}^+_{{\mc{A}}}\cup{{\umode}}^+_{{\mc{B}}}, {{\umode}}^+_{{\mc{A}}}=\overline{{{\umode}}^+_{{\mc{A}}}}, {{\umode}}^+_{{\mc{B}}}=\overline{{{\umode}}^+_{{\mc{B}}}}. \label{Zer_sq_inc}
\end{eqn}
Note that $\#{{\umode}}_{{\mc{I}}}^+=\#{{\umode}}_{{\mc{I}}}^-$, etc. 

\begin{table}\caption{Wiener--Hopf kernel ((2.8)${}^{\dagger}$, (7.8)${}^{\dagger}$) related details for type I}
\begin{center}
\begin{tabular}{|c|c|l|l|l|l|l|l|l|l||}\hline
S.no&strip&${\ms{N}}$ &${\ms{N}}_{\mathfrak{a}}$&${\ms{N}}_{\mathfrak{b}}$&${\ms{D}}$ &Figure\\\hline
\hline
(a) &${\mathfrak{S}\hspace{-.4ex}}\sqkX$&${\mathtt{V}}_{{\mathtt{N}_{\mathfrak{a}}}}{\mathtt{V}}_{{\mathtt{N}_{\mathfrak{b}}}}$&${\mathtt{V}}_{{\mathtt{N}_{\mathfrak{a}}}}$&${\mathtt{V}}_{{\mathtt{N}_{\mathfrak{b}}}}$&${\mathtt{U}}_{{\mathtt{N}}}$&2(a)${}^{\dagger}$\\
(b) &${\mathfrak{S}\hspace{-.4ex}}\sqkR$&${\mathpzc{H}}{\mathtt{U}}_{{\mathtt{N}_{\mathfrak{a}}}-1}{\mathtt{U}}_{{\mathtt{N}_{\mathfrak{b}}}-1}$&${\mathpzc{H}}{\mathtt{U}}_{{\mathtt{N}_{\mathfrak{a}}}-1}$&${\mathpzc{H}}{\mathtt{U}}_{{\mathtt{N}_{\mathfrak{b}}}-1}$&${\mathpzc{H}}{\mathtt{U}}_{{\mathtt{N}}-1}$&
2(b)${}^{\dagger}$\\
(c) &${\mathfrak{S}\hspace{-.4ex}}\sqkXR$&${\mathpzc{H}}{\mathtt{V}}_{{\mathtt{N}_{\mathfrak{a}}}}{\mathtt{U}}_{{\mathtt{N}_{\mathfrak{b}}}-1}$&${\mathtt{V}}_{{\mathtt{N}_{\mathfrak{a}}}}$&${\mathpzc{H}}{\mathtt{U}}_{{\mathtt{N}_{\mathfrak{b}}}-1}$&${\mathtt{V}}_{{\mathtt{N}}}$&
2(c)${}^{\dagger}$\\
\hline
\end{tabular}
\end{center}
\label{bifurcatedstripbc_sq_ND}
\end{table}

\begin{table}
\caption{Wiener--Hopf kernel ((3.4a)${}^{\dagger}$, \eqref{WHCeq_sq_gen_altinc}) related details 
for 
type II}
\begin{center}
\begin{tabular}{|c|c|l|l|l|l|l|l|l|l||}
\hline
S.no&strip&${\ms{N}}$&$\mathring{{\ms{D}}}$&$\mathring{{\ms{D}}}_{\mathfrak{a}}$&$\mathring{{\ms{D}}}_{\mathfrak{b}}$
&Figure\\
\hline
\hline
(a') &${\mathfrak{S}\hspace{-.4ex}}\sqcX$&${\mathtt{U}}_{{\mathtt{N}}}$&${\mathtt{U}}_{{\mathtt{N}_{\mathfrak{a}}}}{\mathtt{U}}_{{\mathtt{N}_{\mathfrak{b}}}}$&${\mathtt{U}}_{{\mathtt{N}_{\mathfrak{a}}}}$&${\mathtt{U}}_{{\mathtt{N}_{\mathfrak{b}}}}$
&2(a')${}^{\dagger}$\\
(b') &${\mathfrak{S}\hspace{-.4ex}}\sqcR$&${\mathpzc{H}}{\mathtt{U}}_{{\mathtt{N}}-1}$&${\mathtt{V}}_{{\mathtt{N}_{\mathfrak{a}}}}{\mathtt{V}}_{{\mathtt{N}_{\mathfrak{b}}}}$&${\mathtt{V}}_{{\mathtt{N}_{\mathfrak{a}}}}$&${\mathtt{V}}_{{\mathtt{N}_{\mathfrak{b}}}}$
&2(b')${}^{\dagger}$\\
(c') &${\mathfrak{S}\hspace{-.4ex}}\sqcXR$&${\mathtt{V}}_{{\mathtt{N}}}$&${\mathtt{U}}_{{\mathtt{N}_{\mathfrak{a}}}}{\mathtt{V}}_{{\mathtt{N}_{\mathfrak{b}}}}$&${\mathtt{U}}_{{\mathtt{N}_{\mathfrak{a}}}}$&${\mathtt{V}}_{{\mathtt{N}_{\mathfrak{b}}}}$
&2(c')${}^{\dagger}$\\
\hline
\end{tabular}
\end{center}
\label{bifurcatedstripbc_sq_ND_c}
\end{table}

Similar to \eqref{farfield_k_sq} and \eqref{farfield_c_sq}, for the incidence from the left {(as remarked before, the relevant details are provided in Appendix \ref{app_exactsol})},
\begin{eqn}
{\mathtt{u}}^{{t}}_{{\mathtt{x}}, {\mathtt{y}}}&\sim{{\mathrm{A}}}{\seigen}_{({{\kappa}^{i}}){{\nu}}}{{z}}_{{P}}^{{\mathtt{x}}}+{{\mathrm{A}}}\frac{{\ms{N}}_-({{z}}_{{P}})}{{\ms{D}}_-({{z}}_{{P}})}\sum\nolimits_{{{\umode}}_{{\mc{I}}}^+}\frac{{{{\mathrm{v}}}_{({{\kappa}^{i}})}}}{{{\mathrm{v}}}_{({\kappa}_{{z}})}}\frac{{\seigen}_{({\kappa}_{{z}}){{\nu}}}{{z}}^{{\mathtt{x}}}}{{{z}}-{{z}}_{{P}}} \frac{{\ms{N}}_+({{z}})}{{\ms{D}}'_+({{z}})}\\
{\mathtt{u}}^{{t}}_{{\mathtt{x}}, {\mathtt{y}}}&\sim{{\mathrm{A}}}\frac{{\ms{N}}_-({{z}}_{{P}})}{{\ms{D}}_-({{z}}_{{P}})}\sum\nolimits_{{{\umode}}^{-}_{{\mc{A}}{\mc{B}}}}\frac{{{{\mathrm{v}}}_{({{\kappa}^{i}})}}}{{{\mathrm{v}}}_{({\kappa}_{{z}})}}\frac{{\seigen}_{({\kappa}_{{z}}){{\nu}}}{{z}}^{{\mathtt{x}}}}{{{z}}-{{z}}_{{P}}} \frac{{\ms{D}}_-({{z}})}{{\ms{N}}'_-({{z}})},
\label{farfield_k_sq_altinc}
\end{eqn}
for 
type I bifurcation, as ${\mathtt{x}}\to+\infty$ and ${\mathtt{x}}\to-\infty$, respectively, and
\begin{eqn}
{\mathtt{u}}^{{t}}_{{\mathtt{x}}, {\mathtt{y}}}&\sim{{\mathrm{A}}}{\seigen}_{({{\kappa}^{i}}){{\nu}}}{{z}}_{{P}}^{{\mathtt{x}}}+{{\mathrm{A}}}\frac{{\ms{D}}_-({{z}}_{{P}})}{{\ms{N}}_-({{z}}_{{P}})}\sum\nolimits_{{{z}}\in{\umode}_{{\mc{I}}}^+}\frac{({\seigen}_{({{\kappa}^{i}}){{\mathtt{N}_{\mathfrak{b}}}+2}}+{\seigen}_{({{\kappa}^{i}}){{\mathtt{N}_{\mathfrak{b}}}}})}{{{\mathpzc{Q}}}({{z}}){\seigen}_{({\kappa}_{{z}}){{\mathtt{N}_{\mathfrak{b}}}+1}}}
\\&
\frac{1}{{{z}}_{{P}}-{{z}}_{{\mathpzc{q}}}^{-1}}\frac{{{z}}-{{z}}_{{\mathpzc{q}}}^{-1}}{{{z}}-{{z}}_{{P}}}\frac{{\ms{D}}_+({{z}})}{{\ms{N}}'_+({{z}})}{\seigen}_{({\kappa}_{{z}}){{\nu}}}{{z}}^{{\mathtt{x}}},\\{\mathtt{u}}^{{t}}_{{\mathtt{x}}, {\mathtt{y}}}&\sim{{\mathrm{A}}}\frac{{\ms{D}}_-({{z}}_{{P}})}{{\ms{N}}_-({{z}}_{{P}})}\sum\nolimits_{{{z}}\in{{\umode}}^{-}_{{\mc{A}}{\mc{B}}}}\frac{({\seigen}_{({{\kappa}^{i}}){{\mathtt{N}_{\mathfrak{b}}}+2}}+{\seigen}_{({{\kappa}^{i}}){{\mathtt{N}_{\mathfrak{b}}}}})}{{\seigen}_{({\kappa}_{{z}}){{\mathtt{N}_{\mathfrak{b}}}}+2}+{\seigen}_{({\kappa}_{{z}}){{\mathtt{N}_{\mathfrak{b}}}}}}
\\&
\frac{1}{{{z}}_{{P}}-{{z}}_{{\mathpzc{q}}}^{-1}}\frac{{{z}}-{{z}}_{{\mathpzc{q}}}^{-1}}{{{z}}-{{z}}_{{P}}}\frac{{\ms{N}}_-({{z}})}{{\ms{D}}'_-({{z}})}{\seigen}_{({\kappa}_{{z}}){{\nu}}}{{z}}^{{\mathtt{x}}},
\label{farfield_c_sq_altinc}
\end{eqn}
for 
type II bifurcation, as ${\mathtt{x}}\to+\infty$ and ${\mathtt{x}}\to-\infty$, respectively.
Above can be identified with \eqref{assumedfarfield_sq_altinc},
assuming ${{z}}_{{P}}$ corresponds to {either of} the index $\tilde{{\mathsf{a}}}$ and $\tilde{{\mathsf{b}}}$.

For each of the wave modes appearing in the expressions \eqref{farfield_k_sq}, \eqref{farfield_c_sq}, and \eqref{farfield_k_sq_altinc}, \eqref{farfield_c_sq_altinc}, the energy flux can be determined as a product of energy density and the group velocity \cite{Brillouin}. The energy flux, 
for instance, for the incident wave mode 
is given by
\begin{eqn}
{\ms{E}^{i}}(\tilde{{\mathsf{i}}})&=\sum\limits_{{\mathtt{y}}\in\Z_0^{{\mathtt{N}}-1}}|{{\mathrm{A}}}|^2|{\seigen}_{({\kappa}^{i}){{\nu}}}|^2|\velG({\upxi}_{{P}})|=\frac{|{{\mathrm{A}}}|^2}{{\upomega}}\sin|{\upxi}_{{P}}|.
\label{energyflux_inc}
\end{eqn}
Due to the simplicity of assumed square lattice structure, ${\upxi}_{{P}}={\upkappa}_{{\mathtt{x}}}>0$ when the wave is incident {from the terminal ${\mc{I}}$}, i.e., the intact portion, while ${\upxi}_{{P}}<0$ when the wave is incident from {either of the two terminals ${\mc{A}}$ or ${\mc{B}}$}, i.e., the bifurcated portions.

The reflectance and transmittance ({same as Eq. (5.3), (5.5), (5.6) and Remark 5.1 of \paperone{}}) 
are given by the following general expressions
for the wave incidence from the ${\mc{I}}$ terminal (assuming ${{z}}_{{P}}$ corresponds to the index ${\tilde{{{\mathsf{i}}}}}$):
\beqan
{\ms{R}}_{{\mc{I}}}({\tilde{{{\mathsf{i}}}}})&=&
{\mathtt{C}}_{RT}({\tilde{{{\mathsf{i}}}}})\sum\nolimits_{{{z}}\in{{\umode}}^+_{{\mc{I}}}}\frac{\overline{{\ms{D}}_-({{z}}){\ms{N}}_+({{z}})}}{{\ms{N}}_-({{z}}){\ms{D}}'_+({{z}})}\frac{{{z}}_{{P}}}{({{z}}-{{z}}_{{P}})^2},\label{Ref_sq_k}\\
{\ms{T}}_{{\mc{A}}}({\tilde{{{\mathsf{i}}}}})&=&
{\mathtt{C}}_{RT}({\tilde{{{\mathsf{i}}}}})\sum\nolimits_{{{z}}\in{{\umode}}^-_{{\mc{A}}}}\frac{\overline{{\ms{D}}_-({{z}}){\ms{N}}_+({{z}})}}{{\ms{N}}'_-({{z}}){\ms{D}}_+({{z}})}\frac{{{z}}_{{P}}}{({{z}}-{{z}}_{{P}})^2},\label{Trans_sq_kup}\\
{\ms{T}}_{{\mc{B}}}({\tilde{{{\mathsf{i}}}}})&=&
{\mathtt{C}}_{RT}({\tilde{{{\mathsf{i}}}}})\sum\nolimits_{{{z}}\in{{\umode}}^-_{{\mc{B}}}}\frac{\overline{{\ms{D}}_-({{z}}){\ms{N}}_+({{z}})}}{{\ms{N}}'_-({{z}}){\ms{D}}_+({{z}})}\frac{{{z}}_{{P}}}{({{z}}-{{z}}_{{P}})^2},\label{Trans_sq_klow}
\eeqan
\begin{eqn}
\text{where }
{\mathtt{C}}_{RT}({\tilde{{{\mathsf{i}}}}})&=\frac{{{z}}_{{P}}{\ms{N}}_-({{z}}_{{P}}){\ms{D}}_+({{z}}_{{P}})}{\overline{{\ms{D}}'_-({{z}}_{{P}})}\overline{{\ms{N}}_+({{z}}_{{P}})}}.
\label{CRexp_k}
\end{eqn}
With ${\ms{D}}_+({{z}})={\mathpzc{Q}}_+({{z}})\mathring{{\ms{D}}}_+({{z}})$ and ${\ms{D}}'_-({{z}})={\mathpzc{Q}}_-({{z}})\mathring{{\ms{D}}}'_-({{z}})$ for the 
type II bifurcation, the same expressions hold (after swapping \cite{Bls9s} ${\ms{N}}$ and ${\ms{D}}$). 
In particular, for the wave incidence from the 
${\mc{I}}$ terminal in the 
type II bifurcation
(assuming ${{z}}_{{P}}$ corresponds to the index ${\tilde{{{\mathsf{i}}}}}$, i.e., ${{z}}_{{P}}={{z}}_{\tilde{{\mathsf{i}}}}$,),
\beqan
{\ms{R}}_{{\mc{I}}}({\tilde{{{\mathsf{i}}}}})&=&
{\mathtt{C}}_{RT}({\tilde{{{\mathsf{i}}}}})\sum\nolimits_{{{z}}\in{{\umode}}^+_{{\mc{I}}}}\frac{\overline{{\ms{N}}_-({{z}})\mathring{{\ms{D}}}_+({{z}})}}{\mathring{{\ms{D}}}_-({{z}}){\ms{N}}'_+({{z}})}\frac{{{z}}_{{P}}}{({{z}}-{{z}}_{{P}})^2},\label{Ref_sq_c}\\
{\ms{T}}_{{\mc{A}}}({\tilde{{{\mathsf{i}}}}})&=&
{\mathtt{C}}_{RT}({\tilde{{{\mathsf{i}}}}})\sum\nolimits_{{{z}}\in{{\umode}}^-_{{\mc{A}}}}\frac{\overline{{\ms{N}}_-({{z}})\mathring{{\ms{D}}}_+({{z}})}}{\mathring{{\ms{D}}}'_-({{z}}){\ms{N}}_+({{z}})}\frac{{{z}}_{{P}}}{({{z}}-{{z}}_{{P}})^2},\label{Trans_sq_cup}\\
{\ms{T}}_{{\mc{B}}}({\tilde{{{\mathsf{i}}}}})&=&
{\mathtt{C}}_{RT}({\tilde{{{\mathsf{i}}}}})\sum\nolimits_{{{z}}\in{{\umode}}^-_{{\mc{B}}}}\frac{\overline{{\ms{N}}_-({{z}})\mathring{{\ms{D}}}_+({{z}})}}{\mathring{{\ms{D}}}'_-({{z}}){\ms{N}}_+({{z}})}\frac{{{z}}_{{P}}}{({{z}}-{{z}}_{{P}})^2},\label{Trans_sq_clow}
\eeqan
\begin{eqn}
\text{where }
{\mathtt{C}}_{RT}({\tilde{{{\mathsf{i}}}}})&=\frac{{{z}}_{{P}}\mathring{{\ms{D}}}_-({{z}}_{{P}}){\ms{N}}_+({{z}}_{{P}})}{\overline{{\ms{N}}'_-({{z}}_{{P}})}\overline{\mathring{{\ms{D}}}_+({{z}}_{{P}})}}.
\label{CRexp_c}
\end{eqn}

In the context of this paper, it is natural curiosity whether the procedure of 
\paperone{} can be also repeated for the incidence from the bifurcated portions and recover the expressions same as or similar to \eqref{Ref_sq_c}, \eqref{Trans_sq_cup}, \eqref{Trans_sq_clow} and \eqref{CRexp_c}. 
Indeed, for the wave incidence from {the terminal} ${\mc{A}}$ or ${\mc{B}}$, while omitting details as they are similar to the case of incidence from terminal $\mc{I}$ provided in \paperone, it is found that the reflectance and transmittance are given by 
(assuming ${{z}}_{{P}}$ corresponds to the index ${\tilde{{\mathsf{a}}}}$, i.e., ${{z}}_{{P}}={{z}}_{\tilde{{\mathsf{a}}}}$, for \eqref{RTupinc_sq_altinc1}, \eqref{RTupinc_sq_altinc2}, and to ${\tilde{{\mathsf{b}}}}$, i.e., ${{z}}_{{P}}={{z}}_{\tilde{{\mathsf{b}}}}$, for \eqref{RTlowinc_sq_altinc1}, \eqref{RTlowinc_sq_altinc2})
\beqan
{\ms{R}}_{{\mc{A}}}({\tilde{{\mathsf{a}}}})&=&
{\mathtt{C}}_{RT}({\tilde{{\mathsf{a}}}})\sum\nolimits_{{{z}}\in{{\umode}}^-_{{\mc{A}}}}\frac{\overline{{\ms{D}}_-({{z}}){\ms{N}}_+({{z}})}}{{\ms{N}}'_-({{z}}){\ms{D}}_+({{z}})}\frac{{{z}}_{{P}}}{({{z}}-{{z}}_{{P}})^2}, \label{RTupinc_sq_altinc1}\\
{\ms{T}}_{{\mc{B}}}({\tilde{{\mathsf{a}}}})&=&
{\mathtt{C}}_{RT}({\tilde{{\mathsf{a}}}})\sum\nolimits_{{{z}}\in{{\umode}}^+_{{\mc{B}}}}\frac{\overline{{\ms{D}}_-({{z}}){\ms{N}}_+({{z}})}}{{\ms{N}}_-({{z}}){\ms{D}}'_+({{z}})}\frac{{{z}}_{{P}}}{({{z}}-{{z}}_{{P}})^2},\label{RTupinc_sq_altinc2}
\eeqan
\beqan
{\ms{R}}_{{\mc{B}}}({\tilde{{\mathsf{b}}}})&=&
{\mathtt{C}}_{RT}({\tilde{{\mathsf{b}}}})\sum\nolimits_{{{z}}\in{{\umode}}^-_{{\mc{B}}}}\frac{\overline{{\ms{D}}_-({{z}}){\ms{N}}_+({{z}})}}{{\ms{N}}'_-({{z}}){\ms{D}}_+({{z}})}\frac{{{z}}_{{P}}}{({{z}}-{{z}}_{{P}})^2}, \label{RTlowinc_sq_altinc1}\\
{\ms{T}}_{{\mc{A}}}({\tilde{{\mathsf{b}}}})&=&
{\mathtt{C}}_{RT}({\tilde{{\mathsf{b}}}})\sum\nolimits_{{{z}}\in{{\umode}}^+_{{\mc{A}}}}\frac{\overline{{\ms{D}}_-({{z}}){\ms{N}}_+({{z}})}}{{\ms{N}}_-({{z}}){\ms{D}}'_+({{z}})}\frac{{{z}}_{{P}}}{({{z}}-{{z}}_{{P}})^2},\label{RTlowinc_sq_altinc2}
\eeqan
and the common expression for the transmittance and ${\mathtt{C}}_{RT}$ are
\begin{eqn}
{\ms{T}}_{{\mc{I}}}({\tilde{{\mathsf{a}}}})={\ms{T}}_{{\mc{I}}}({\tilde{{\mathsf{b}}}})&=
{\mathtt{C}}_{RT}\sum\nolimits_{{{z}}\in{{\umode}}^+_{{\mc{I}}}}\frac{\overline{{\ms{D}}_-({{z}}){\ms{N}}_+({{z}})}}{{\ms{N}}_-({{z}}){\ms{D}}'_+({{z}})}\frac{{{z}}_{{P}}}{({{z}}-{{z}}_{{P}})^2},
\label{Trans_sq_altinc}
\end{eqn}
\begin{eqn}
\text{where }
{\mathtt{C}}_{RT}={\mathtt{C}}_{RT}({\tilde{{\mathsf{a}}}})={\mathtt{C}}_{RT}({\tilde{{\mathsf{b}}}})&=\frac{{{z}}_{{P}}{\ms{N}}_-({{z}}_{{P}}){\ms{D}}_+({{z}}_{{P}})}{\overline{{\ms{N}}'_+({{z}}_{{P}})}\overline{{\ms{D}}_-({{z}}_{{P}})}},
\label{CRexp_altinc}
\end{eqn}
where ${{z}}_{{P}}$ corresponds to ${\tilde{{\mathsf{a}}}}$ and ${\tilde{{\mathsf{b}}}}$ for ${\mathtt{C}}_{RT}({\tilde{{\mathsf{a}}}})$ and ${\mathtt{C}}_{RT}({\tilde{{\mathsf{b}}}})$, respectively.
Analogous expressions hold for the incidence from the bifurcated portion of the 
type II bifurcation, taking cue from the {sentence} preceding {that containing} \eqref{Ref_sq_c}, \eqref{Trans_sq_cup}, \eqref{Trans_sq_clow}, and \eqref{CRexp_c}. 

\section{Conductance matrix}
{According to Fig. \ref{latticestrip_sq_BCs_asymm_semiinfdefect}},
the elements of ${{\umode}}^+_{{\mc{I}}}$ (resp. ${{\umode}}^-_{{\mc{I}}}$) are indexed
by ${{\mathsf{i}}}$ (resp. ${\tilde{{{\mathsf{i}}}}}$) with a range $1\dotsc N^{{\mc{I}}}=\#{{\umode}}^+_{{\mc{I}}}$, while the elements of ${{\umode}}^-_{{\mc{A}},{\mc{B}}}$ (resp. ${{\umode}}^+_{{\mc{A}},{\mc{B}}}$) are indexed by ${{{\mathsf{a}}},{{\mathsf{b}}}}$ (resp. ${\tilde{{\mathsf{a}}}},{\tilde{{\mathsf{b}}}}$) ranging from $1$ to $N^{{{\mc{A}},{\mc{B}}}}=\#{{\umode}}^-_{{\mc{A}},{\mc{B}}}$ (note that $N^{{{\mc{A}}}}\le{\mathtt{N}_{\mathfrak{a}}}$ and $N^{{{\mc{A}}}}\le{\mathtt{N}_{\mathfrak{b}}}$ due to assumed lattice structure). 
With these symbols for channels,
the Landauer--B{\"{u}}ttiker formalism \cite{Landauer1957,Landauer,Buttiker1986} 
relates the scattering matrix to the conductance matrix components of the sample (with three terminals) as
\begin{eqn}
{\mathscr{G}}_{{{\mc{A}}}{\mc{I}}}
={\mathscr{G}}_{{\mc{I}}{{\mc{A}}}}
 &{:=} 
\text{Tr}({{\tau}}^{{\mc{I}}{\tilde{{\mc{A}}}}}_{N^{{\mc{I}}}\times N^{{\mc{A}}}}{{{\tau}}^{{\mc{I}}{\tilde{{\mc{A}}}}\dagger}_{N^{{\mc{I}}}\times N^{{\mc{A}}}}}),\\
{\mathscr{G}}_{{{\mc{B}}}{\mc{I}}}
={\mathscr{G}}_{{\mc{I}}{{\mc{B}}}}
 &{:=} 
\text{Tr}({{\tau}}^{{\mc{I}}{\tilde{{\mc{B}}}}}_{N^{{\mc{I}}}\times N^{{\mc{B}}}}{{{\tau}}^{{\mc{I}}{\tilde{{\mc{B}}}}\dagger}_{N^{{\mc{I}}}\times N^{{\mc{B}}}}}),\\
{\mathscr{G}}_{{{\mc{A}}}{{\mc{B}}}}
={\mathscr{G}}_{{{\mc{B}}}{{\mc{A}}}}
 &{:=} 
\text{Tr}({{\tau}}^{{{\mc{B}}}{\tilde{{\mc{A}}}}}_{N^{{{\mc{B}}}}\times N^{{\mc{A}}}}{{{\tau}}^{{{\mc{B}}}{\tilde{{\mc{A}}}}\dagger}_{N^{{{\mc{B}}}}\times N^{{\mc{A}}}}}).
\label{WSNEeq30}
\end{eqn}
{Note that} the factor ${{z}}_{{P}}/({{z}}-{{z}}_{{P}})^2$ in \eqref{Ref_sq_k}--\eqref{CRexp_altinc} can be also replaced by an alternate expression $-\overline{{z}}/({{z}}-{{z}}_{{P}})\overline{({{z}}-{{z}}_{{P}})}$ {as ${z}, {z}_P$ lie on the unit circle in complex plane}.
The detailed expressions \eqref{WSNEeq30} {can be seen to possess
the general form}
\begin{eqn}
{\mathscr{G}}_{{A}{B}}
=\sum\nolimits_{\tilde{{{A}}}=1}^{N^{{A}}}\sum\nolimits_{{B}=1}^{{N^{{B}}}}\frac{\mc{G}({{z}}_{\tilde{{{A}}}})}{\mc{F}({{z}}_{\tilde{{{A}}}})}\frac{\mc{F}({{z}}_{{B}})}{\mc{G}({{z}}_{{B}})},
\label{genGexp}
\end{eqn}
where ${A},{B}$ belong to the set $\{{\mc{I}}, {{\mc{A}}}, {{\mc{B}}}\}$ and $\mc{F}={{\ms{D}}_-{\ms{N}}_+}$ and $\mc{G}={{\ms{D}}_+{\ms{N}}_-}$ for 
type I bifurcation and with $\mc{F}={{\ms{N}}_-({{z}}_{{\mathsf{a}}})\mathring{{\ms{D}}}_+({{z}}_{{\mathsf{a}}})}$ and $\mc{G}={{\ms{N}}_+({{z}}_{{\mathsf{a}}})\mathring{{\ms{D}}}_-({{z}}_{{\mathsf{a}}})}$ for type II bifurcation.

\begin{figure}[h]
\centering
{\includegraphics[width=.9\linewidth]{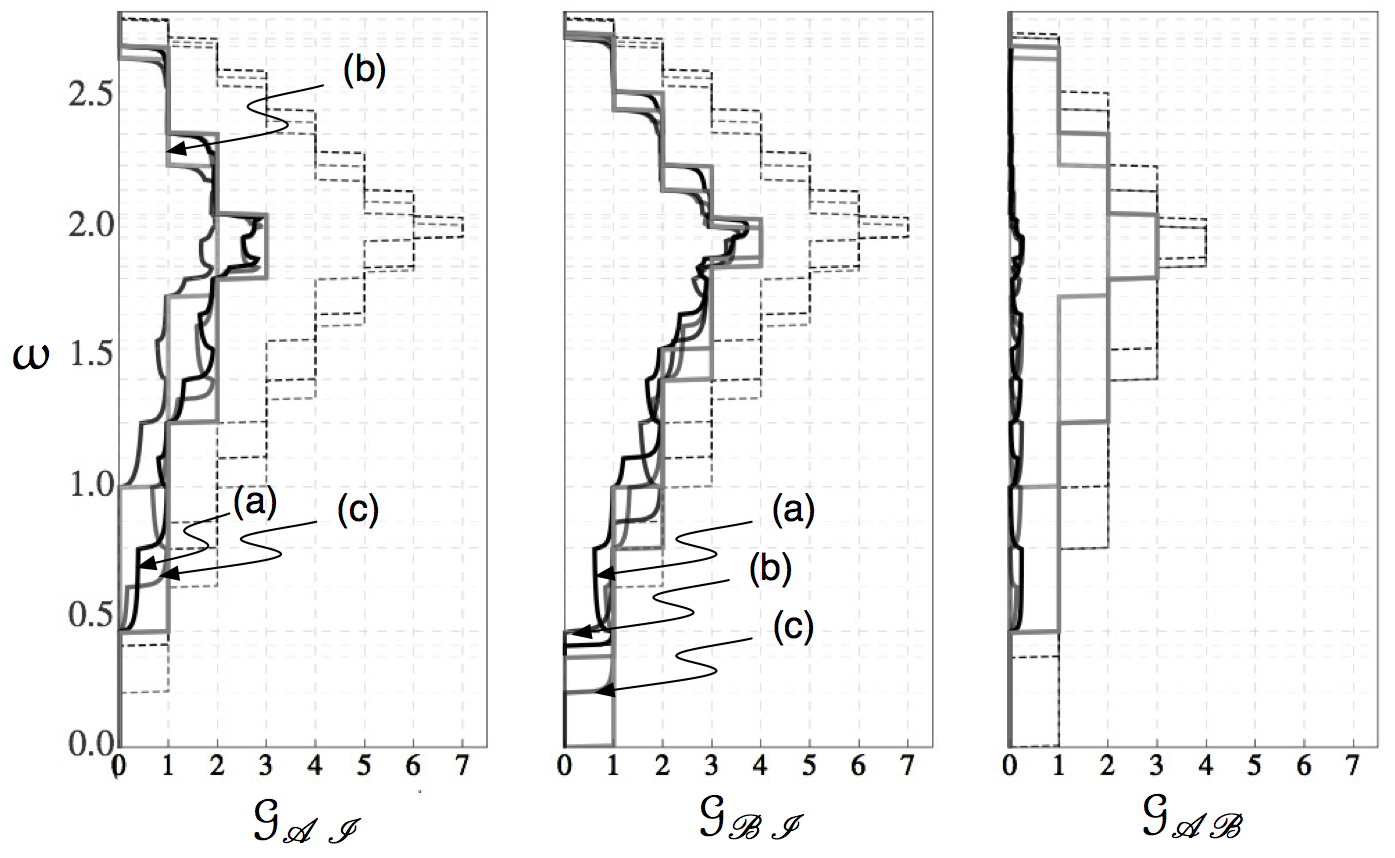}}
\caption{(left to right) 
The conductance matrix elements 
for all three 
bifurcated waveguides of type I 
(${\mathtt{N}_{\mathfrak{a}}}=3, {\mathtt{N}_{\mathfrak{b}}}=4, {\mathtt{N}}=7$), i.e., (a), (b), (c) on the left side of Fig. \ref{latticestrip_sq_BCs_asymm_semiinfdefect}.
}
\label{latticestrip_Conductance_sq_X_R_XR_smallN}
\end{figure}

\begin{figure}[h]
\centering
{\includegraphics[width=.9\linewidth]{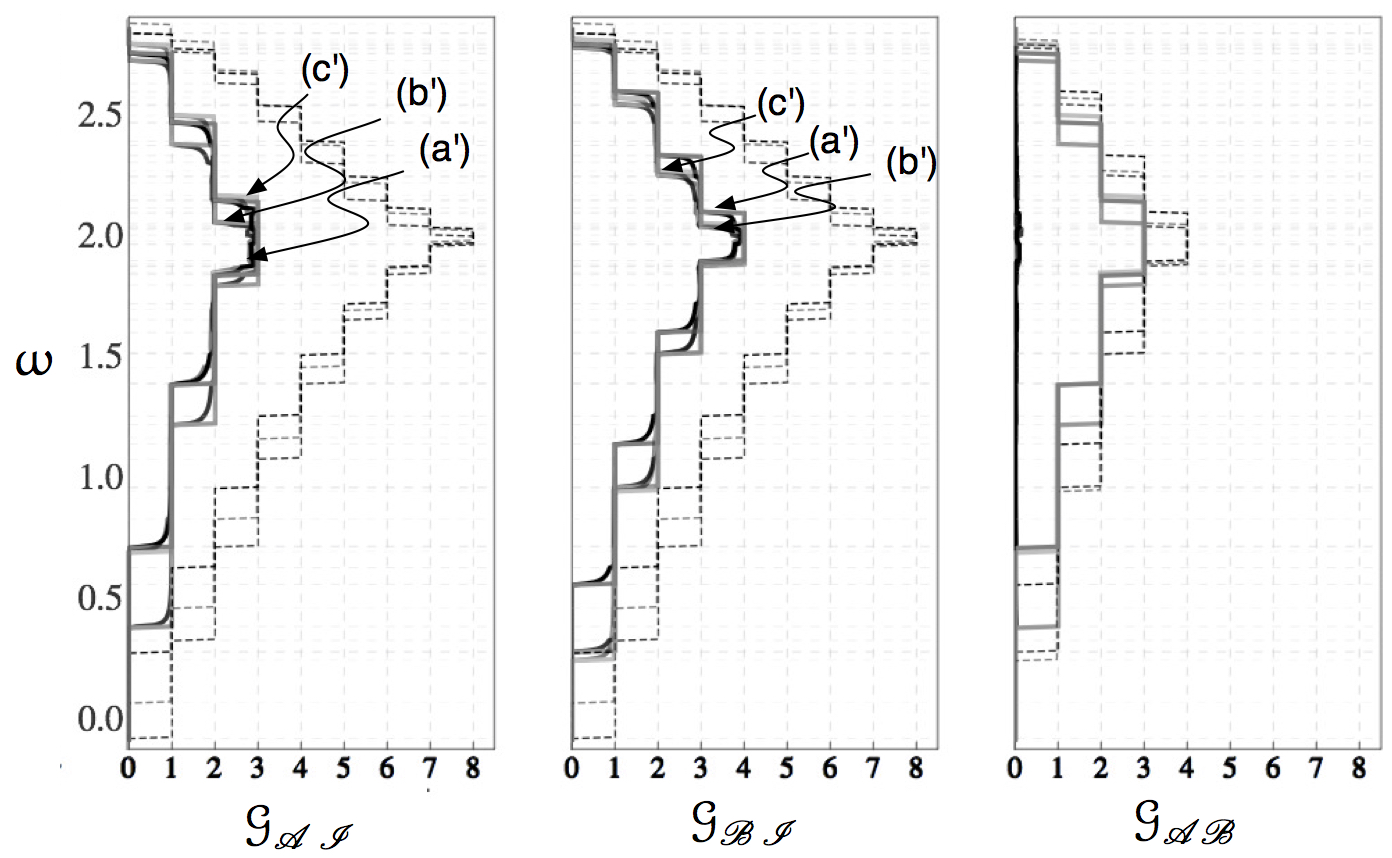}}
\caption{(left to right) 
The conductance matrix elements 
for all three 
bifurcated waveguides of type II 
(${\mathtt{N}_{\mathfrak{a}}}=3, {\mathtt{N}_{\mathfrak{b}}}=4, {\mathtt{N}}=8$), i.e., (a'), (b'), (c') on the right side of Fig. \ref{latticestrip_sq_BCs_asymm_semiinfdefect}.}
\label{latticestrip_Conductance_sq_c_X_R_XR_smallN}
\end{figure}

\begin{figure}[h]
\centering
{\includegraphics[width=.8\linewidth]{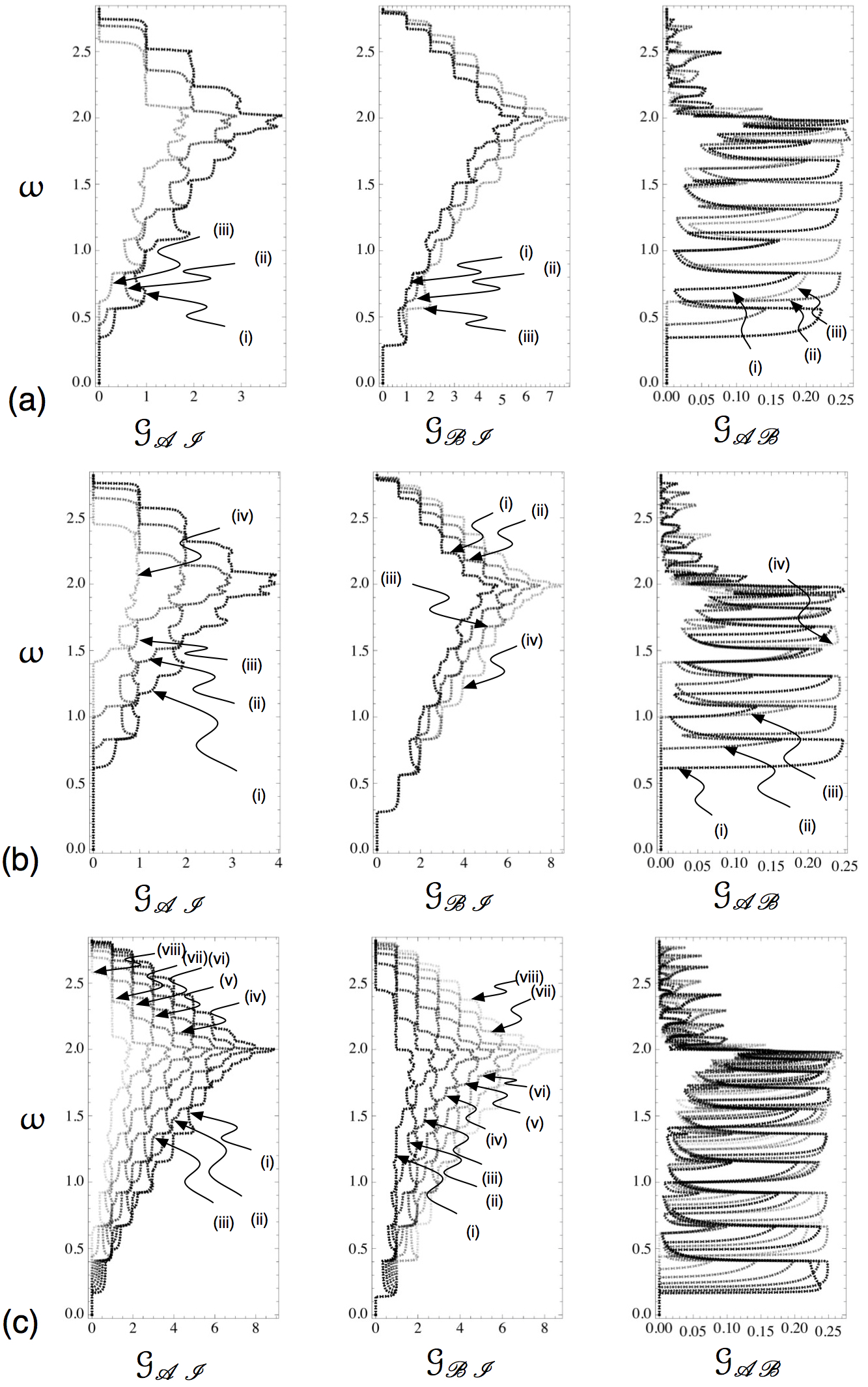}}
\caption{The conductance matrix elements for the type I bifurcated waveguide (a) (${\mathtt{N}}=10$ and ${\mathtt{N}_{\mathfrak{a}}}=2, 3, 4, {\mathtt{N}_{\mathfrak{b}}}={\mathtt{N}}-{\mathtt{N}_{\mathfrak{a}}}$ corresponding to (i)--(iii)), 
(b) (${\mathtt{N}}=11$ and ${\mathtt{N}_{\mathfrak{a}}}=2, 3, 4, 5, {\mathtt{N}_{\mathfrak{b}}}={\mathtt{N}}-{\mathtt{N}_{\mathfrak{a}}}$ corresponding to (i)--(iv)), 
(c) (${\mathtt{N}}=11$ and ${\mathtt{N}_{\mathfrak{a}}}=2, 3, 4, 5, 6, 7, 8, 9, {\mathtt{N}_{\mathfrak{b}}}={\mathtt{N}}-{\mathtt{N}_{\mathfrak{a}}}$ corresponding to (i)--(viii)) listed as (a), (b), (c), respectively, in Table \ref{bifurcatedstripbc_sq_ND} and Fig. \ref{latticestrip_sq_BCs_asymm_semiinfdefect}.}
\label{latticestrip_Conductance_sq_X_R_XR_Nvar}
\end{figure}

\begin{figure}[h]
\centering
{\includegraphics[width=.8\linewidth]{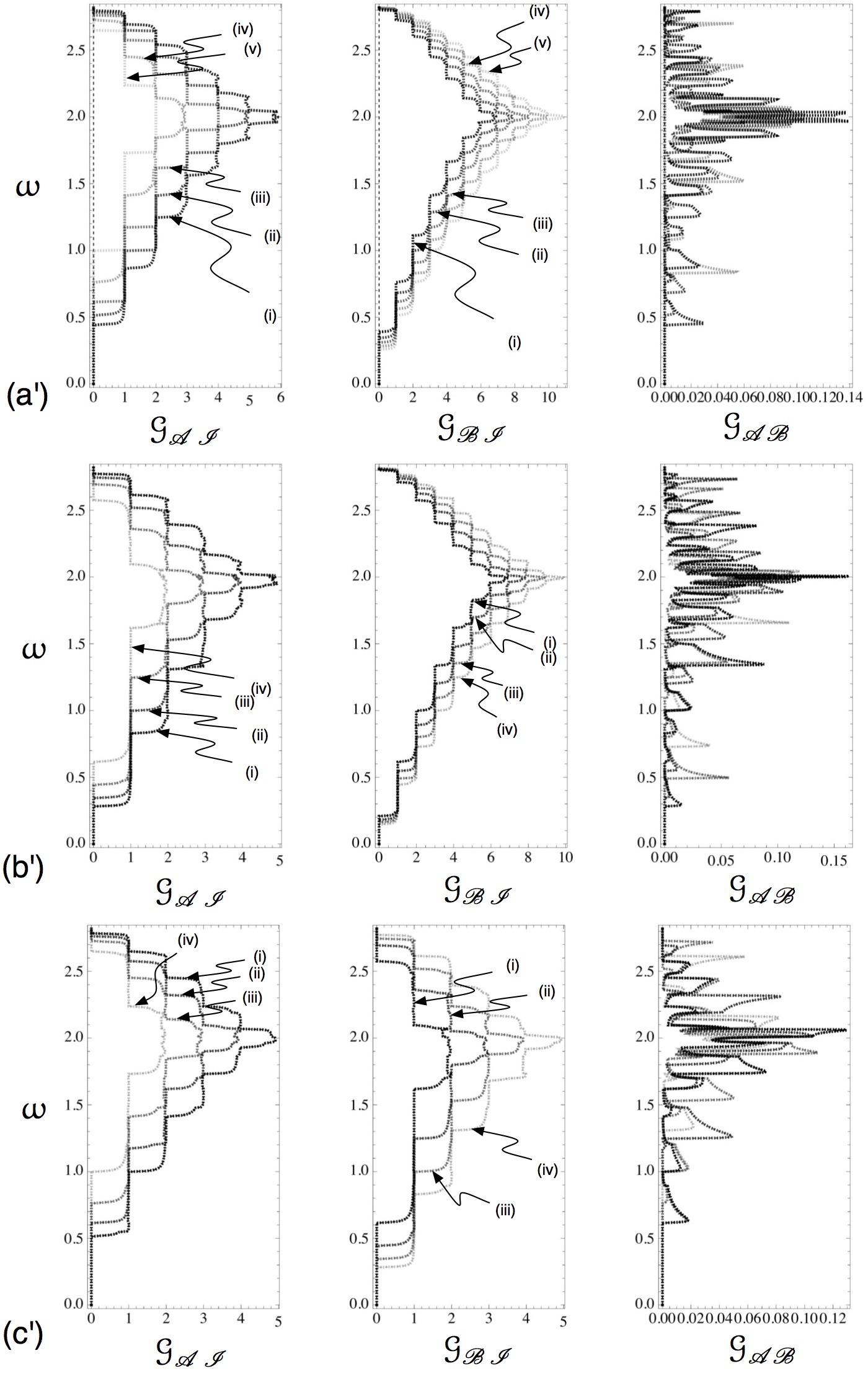}}
\caption{The conductance matrix elements for the type II bifurcated waveguide 
(a') 
(${\mathtt{N}}=14$ and ${\mathtt{N}_{\mathfrak{a}}}=2, 3, 4, 5, 6, {\mathtt{N}_{\mathfrak{b}}}={\mathtt{N}}-1-{\mathtt{N}_{\mathfrak{a}}}$ corresponding to (i)--(v)), 
(b') 
(${\mathtt{N}}=13$ and ${\mathtt{N}_{\mathfrak{a}}}=2, 3, 4, 5, {\mathtt{N}_{\mathfrak{b}}}={\mathtt{N}}-1-{\mathtt{N}_{\mathfrak{a}}}$ corresponding to (i)--(iv)), 
(c') 
(${\mathtt{N}}=8$ and ${\mathtt{N}_{\mathfrak{a}}}=2, 3, 4, 5, {\mathtt{N}_{\mathfrak{b}}}={\mathtt{N}}-1-{\mathtt{N}_{\mathfrak{a}}}$ corresponding to (i)--(iv)) listed as (a'), (b'), (c'), respectively, in Table \ref{bifurcatedstripbc_sq_ND_c} and Fig. \ref{latticestrip_sq_BCs_asymm_semiinfdefect}.}
\label{latticestrip_Conductance_sq_c_X_R_XR_Nvar}
\end{figure}

In fact,
\begin{eqn}
{\mathscr{G}}_{{\mc{I}}{\mc{A}}}
={\mathscr{G}}_{{\mc{A}}{\mc{I}}}
=\sum\nolimits_{\tilde{{\mathsf{i}}}=1}^{N^{{\mc{I}}}}
{\ms{T}}_{{\mc{A}}}(\tilde{{\mathsf{i}}})=\sum\nolimits_{\tilde{{\mathsf{a}}}=1}^{{N^{{\mc{A}}}}}
{\ms{T}}_{{\mc{I}}}(\tilde{{\mathsf{a}}}),\\
{\mathscr{G}}_{{\mc{I}}{\mc{B}}}
={\mathscr{G}}_{{\mc{B}}{\mc{I}}}
=\sum\nolimits_{\tilde{{\mathsf{i}}}=1}^{N^{{\mc{I}}}}
{\ms{T}}_{{\mc{B}}}(\tilde{{\mathsf{i}}})=\sum\nolimits_{\tilde{{\mathsf{b}}}=1}^{{N^{{\mc{B}}}}}
{\ms{T}}_{{\mc{I}}}(\tilde{{\mathsf{b}}}),\\
{\mathscr{G}}_{{\mc{B}}{\mc{A}}}
={\mathscr{G}}_{{\mc{A}}{\mc{B}}}
=\sum\nolimits_{\tilde{{\mathsf{b}}}=1}^{{N^{{\mc{B}}}}}
{\ms{T}}_{{\mc{A}}}(\tilde{{\mathsf{b}}})
=\sum\nolimits_{\tilde{{\mathsf{a}}}=1}^{N^{{\mc{A}}}}
{\ms{T}}_{{\mc{B}}}(\tilde{{\mathsf{a}}}).
\end{eqn}

A graphical illustration of the three off-diagonal components of the conductance $\Ttt{G}$ are provided in Fig. \ref{latticestrip_Conductance_sq_X_R_XR_smallN} and Fig. \ref{latticestrip_Conductance_sq_c_X_R_XR_smallN}.
Labels (a, b, c) and (a${}^\prime$, b${}^\prime$, c${}^\prime$) correspond to boundary 
depicted in Fig. \ref{latticestrip_sq_BCs_asymm_semiinfdefect}. 
{A sample of numerical study accommodating variations in the width of terminals} is presented in Fig. \ref{latticestrip_Conductance_sq_X_R_XR_Nvar} and Fig. \ref{latticestrip_Conductance_sq_c_X_R_XR_Nvar}.
{It is worthy of note that} due to translation symmetry in ${\mathfrak{S}\hspace{-.4ex}}\sqkR$, i.e. case (b) of Fig. \ref{latticestrip_sq_BCs_asymm_semiinfdefect}, there exists a wave mode corresponding to the common lowest branch in the dispersion relations of three waveguides which is transmitted without any change. The corresponding contribution to the conductance is excluded in Fig. \ref{latticestrip_Conductance_sq_X_R_XR_smallN}(b) and Fig. \ref{latticestrip_Conductance_sq_X_R_XR_Nvar}(b).

\begin{figure*}[ht!]
\centering
{\includegraphics[width=\linewidth]{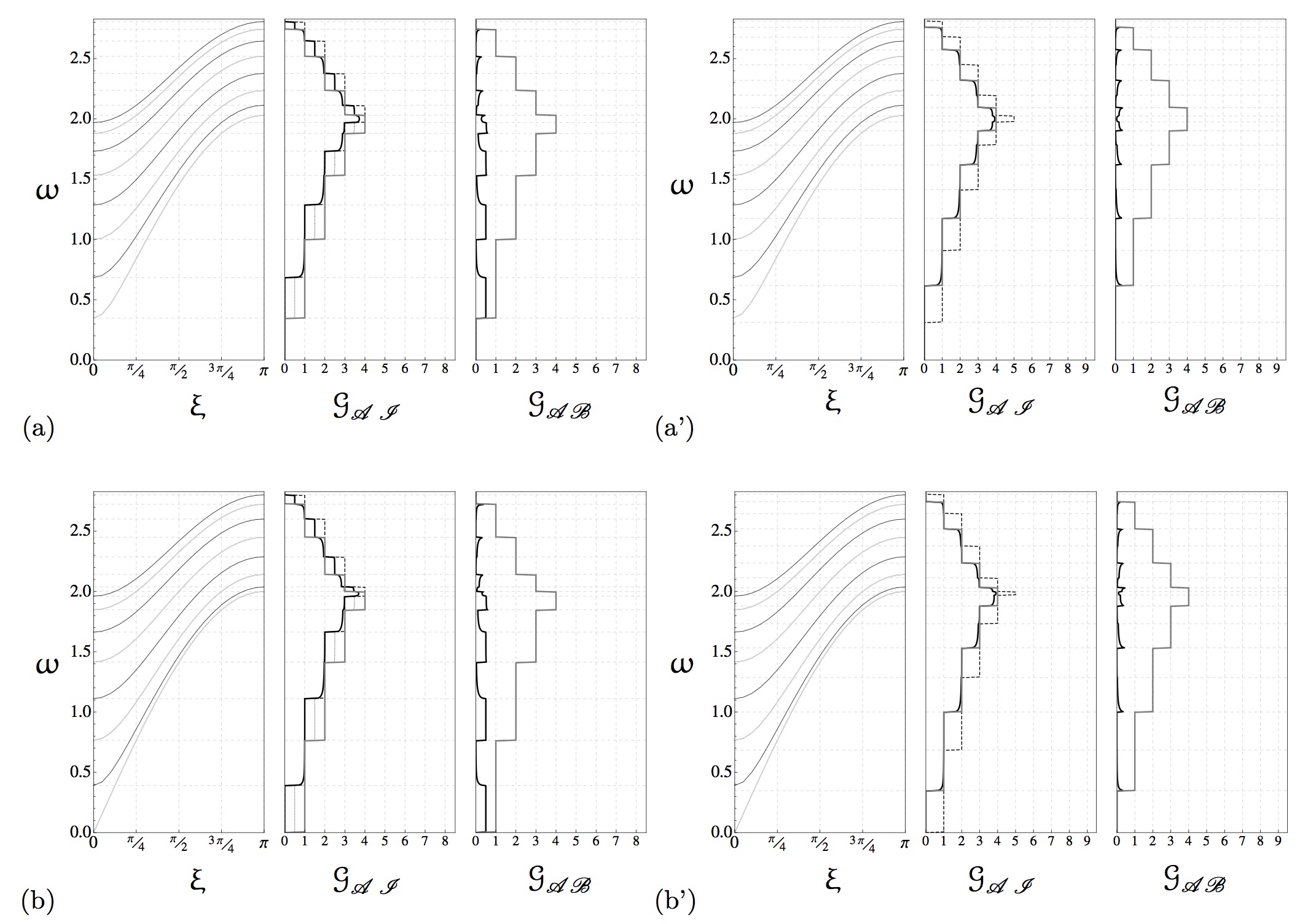}}
\caption{Dispersion curves and the conductance matrix elements ${\mathscr{G}}_{{{\mc{A}}}{\mc{I}}}$ ($={\mathscr{G}}_{{{\mc{B}}}{\mc{I}}}$), ${\mathscr{G}}_{{{\mc{A}}}{\mc{B}}}$ (left to right in each set) for both ((a) and (b)) symmetric type I bifurcated waveguides 
(${\mathtt{N}_{\mathfrak{a}}}={\mathtt{N}_{\mathfrak{b}}}={\mathtt{N}_{\mathfrak{a}}}l=4, {\mathtt{N}}=2{\mathtt{N}_{\mathfrak{a}}}l=8$) as well as for both ((a') and (b')) symmetric type II bifurcated waveguides 
(${\mathtt{N}_{\mathfrak{a}}}={\mathtt{N}_{\mathfrak{b}}}={\mathtt{N}_{\mathfrak{a}}}l=4, {\mathtt{N}}=2{\mathtt{N}_{\mathfrak{a}}}l+1=9$).}
\label{latticestrip_Conductance_sq_symm}
\end{figure*}

\begin{figure}[h]
\centering
{\includegraphics[width=.8\linewidth]{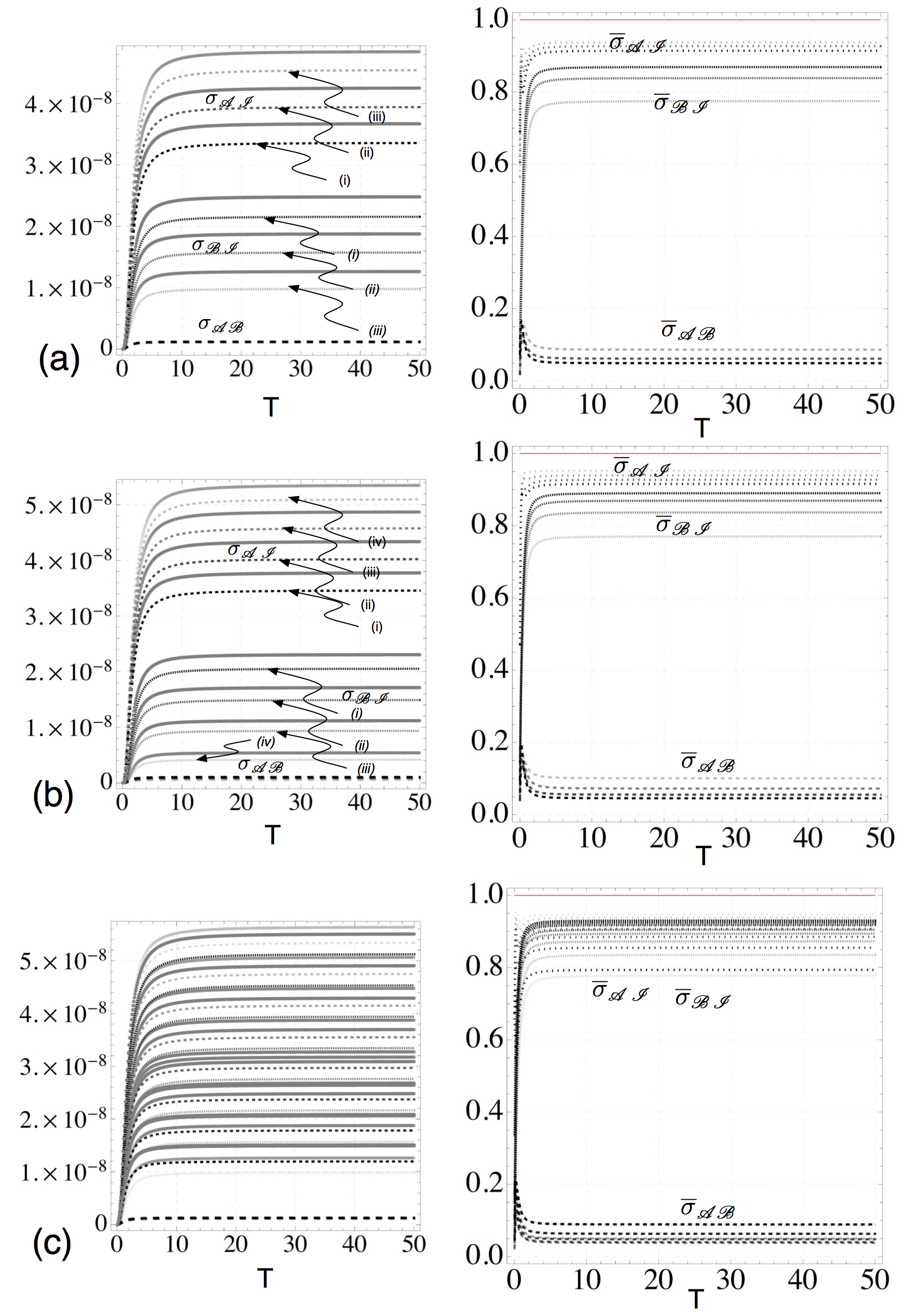}}
\caption{
Thermal conductance matrix elements 
${\sigma}_{\beta\alpha}$ and $\overline{{\sigma}}_{\beta\alpha}$
corresponding to Fig. \ref{latticestrip_Conductance_sq_X_R_XR_Nvar}.
}
\label{latticestrip_ThermConductRelative_sq_XRXR}
\end{figure}

\begin{figure}[h]
\centering
{{\includegraphics[width=.8\linewidth]{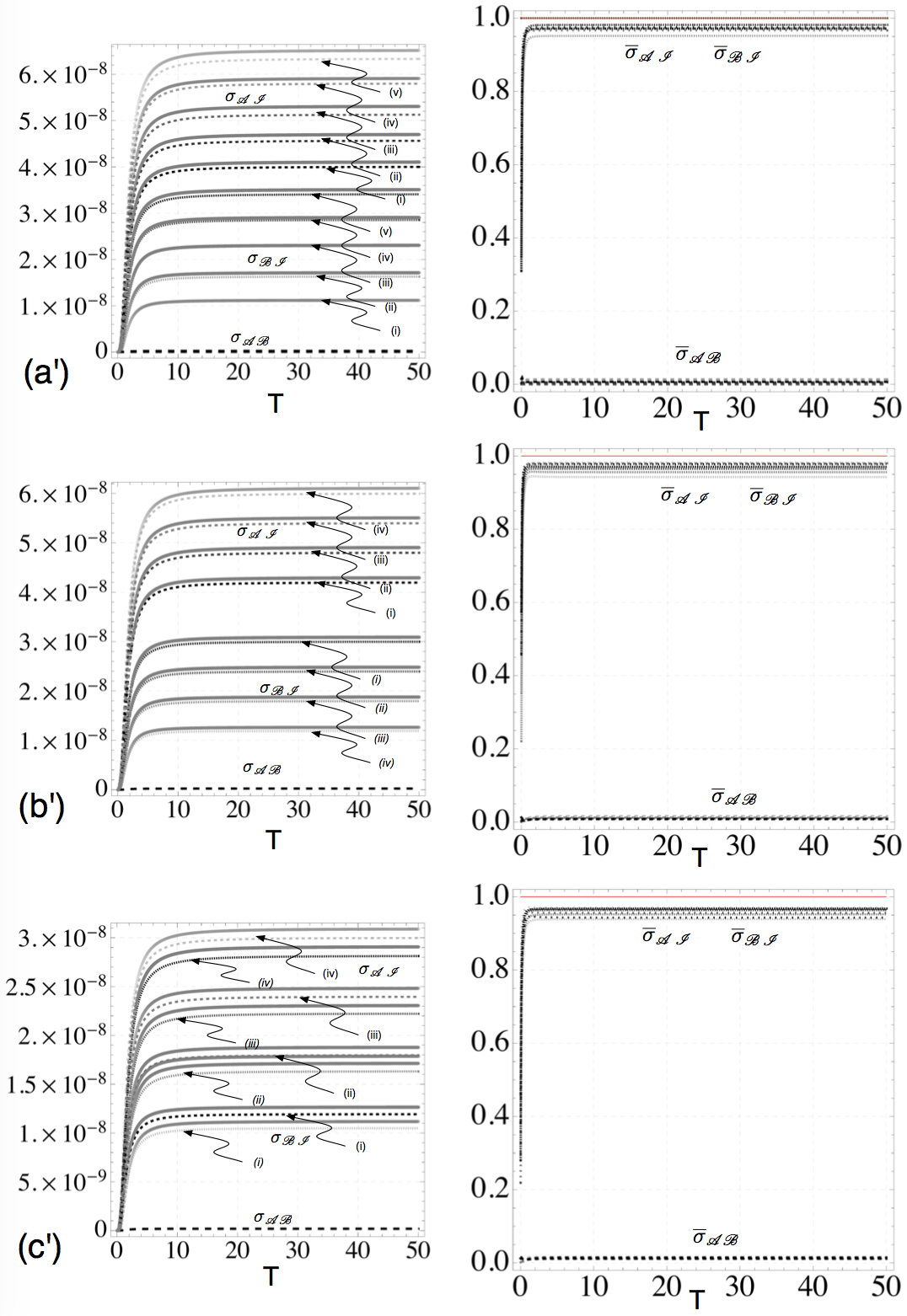}}}
\caption{
Thermal conductance matrix elements 
${\sigma}_{\beta\alpha}$ and $\overline{{\sigma}}_{\beta\alpha}$
corresponding to Fig. \ref{latticestrip_Conductance_sq_c_X_R_XR_Nvar}.
}
\label{latticestrip_ThermConductRelative_sq_c_XRXR}
\end{figure}

\section{Discussion}

\subsection{Symmetric bifurcation} 
\begin{table}
\caption{Wiener--Hopf kernel related details 
for the symmetrically bifurcated square lattice waveguides ($\ms{C}$: common factor)}
\begin{center}
\begin{tabular}{|c|c|l|l|l|l|l||l|}
\hline
S.no&strip&${\ms{N}}$&${\ms{D}}$ or $\mathring{{\ms{D}}}$&W-H eq.
&Figure&$\ms{C}$\\
\hline
\hline
(a) &${\mathfrak{S}\hspace{-.4ex}}\sqkX$&${\mathtt{V}}_{{\mathtt{N}_{\mathfrak{a}}}l}$&${\mathtt{W}}_{{\mathtt{N}_{\mathfrak{a}}}l}$&(2.8)${}^{\dagger}$, (7.8)${}^{\dagger}$&2(a)${}^{\dagger}$&${\mathtt{V}}_{{\mathtt{N}_{\mathfrak{a}}}l}$\\
(b) &${\mathfrak{S}\hspace{-.4ex}}\sqkR$&${\mathpzc{H}}{\mathtt{U}}_{{\mathtt{N}_{\mathfrak{a}}}l-1}$&$2{\mathtt{T}}_{{\mathtt{N}_{\mathfrak{a}}}l}$&(2.8)${}^{\dagger}$, (7.8)${}^{\dagger}$&2(b)${}^{\dagger}$&${\mathpzc{H}}{\mathtt{U}}_{{\mathtt{N}_{\mathfrak{a}}}l-1}$\\
\hline
\hline
(a') &${\mathfrak{S}\hspace{-.4ex}}\sqcX$&$2{\mathtt{T}}_{{\mathtt{N}_{\mathfrak{a}}}l+1}$&$
{\mathtt{U}}_{{\mathtt{N}_{\mathfrak{a}}}l}$&(3.4a)${}^{\dagger}$, \eqref{WHCeq_sq_gen_altinc}&2(a')${}^{\dagger}$&${\mathtt{U}}_{{\mathtt{N}_{\mathfrak{a}}}l}$\\
(b') &${\mathfrak{S}\hspace{-.4ex}}\sqcR$&${\mathpzc{H}}{\mathtt{W}}_{{\mathtt{N}_{\mathfrak{a}}}l}$&$
{\mathtt{V}}_{{\mathtt{N}_{\mathfrak{a}}}l}$&(3.4a)${}^{\dagger}$, \eqref{WHCeq_sq_gen_altinc}&2(b')${}^{\dagger}$&${\mathtt{V}}_{{\mathtt{N}_{\mathfrak{a}}}l}$\\
\hline
\end{tabular}
\end{center}
\label{bifurcatedstripbc_sq_ND_symm}
\end{table}

For the symmetric bifurcated waveguides, i.e. when ${\mathtt{N}_{\mathfrak{a}}}={\mathtt{N}_{\mathfrak{b}}}$, it is easy to see that
${\mathscr{G}}_{{\mc{I}}{\mc{A}}}
={\mathscr{G}}_{{\mc{I}}{\mc{B}}}
$ since ${\ms{T}}_{{\mc{A}}}(\tilde{{\mathsf{i}}})={\ms{T}}_{{\mc{B}}}(\tilde{{\mathsf{i}}})$ so that
${\mathscr{G}}_{{\mc{I}}{\mc{A}}}
={\mathscr{G}}_{{\mc{A}}{\mc{I}}}
={\mathscr{G}}_{{\mc{I}}{\mc{B}}}
={\mathscr{G}}_{{\mc{B}}{\mc{I}}}
={\frac{1}{2}}\sum\nolimits_{\tilde{{\mathsf{i}}}=1}^{N^{{\mc{I}}}}(1-{\ms{R}}_{{\mc{I}}}(\tilde{{\mathsf{i}}})).$
In view of the symmetry, there is an {\em overlapping} property as mentioned in \S{7.2} of 
\paperone{}, the Wiener--Hopf kernal also simplifies (see the kernel related information in Table \ref{bifurcatedstripbc_sq_ND_symm}), hence, the conductance can be written as a sum of two terms one that comes from perfectly transmitted modes (which also correspond to the zeros of the common factor in Table \ref{bifurcatedstripbc_sq_ND_symm} that lie on the unit circle in complex plane) while the other from the scattered modes. Thus, for example, in case of type I bifurcation
\begin{eqn}
{\mathscr{G}}_{{\mc{I}}{\mc{A}}}
={\frac{1}{2}}\sum\limits_\text{{even modes}}1+
{\frac{1}{2}}\sum\nolimits_{\tilde{{{\mathsf{i}}}}=1}^{N^{{\mc{I}}}}\frac{{\ms{D}}_-({{z}}_{\tilde{{{\mathsf{i}}}}}){\ms{N}}_+({{z}}_{\tilde{{{\mathsf{i}}}}})}{\overline{{\ms{N}}'_-({{z}}_{\tilde{{{\mathsf{i}}}}}){\ms{D}}_+({{z}}_{\tilde{{{\mathsf{i}}}}})}}
\sum\nolimits_{{\mathsf{a}}=1}^{{N^{{\mc{A}}}}}\frac{\overline{{\ms{N}}_-({{z}}_{{\mathsf{a}}}){\ms{D}}_+({{z}}_{{\mathsf{a}}})}}{{\ms{D}}'_-({{z}}_{{\mathsf{a}}}){\ms{N}}_+({{z}}_{{\mathsf{a}}})}\frac{{{z}}_{\tilde{{{\mathsf{i}}}}}\overline{{z}}_{{\mathsf{a}}}}{|{{z}}_{{\mathsf{a}}}-{{z}}_{\tilde{{{\mathsf{i}}}}}|^2},
\label{def_g_IA_symm}
\end{eqn}
where ${\ms{N}}$ and ${\ms{D}}$ correspond to Table \ref{bifurcatedstripbc_sq_ND_symm} for (a) and (b). 
Further, when the incidence is from any of the other two terminals, 
say ${\mc{A}}$, 
then
${\ms{R}}_{{\mc{A}}}(\tilde{{\mathsf{a}}})={\ms{T}}_{{\mc{B}}}(\tilde{{\mathsf{a}}}),$
so that
\begin{eqn}
{\mathscr{G}}_{{\mc{A}}{\mc{B}}}
&={\mathscr{G}}_{{\mc{B}}{\mc{A}}}
=-{\frac{1}{2}}\sum\nolimits_{\tilde{{\mathsf{a}}}=1}^{N^{{\mc{A}}}}\frac{{\ms{N}}_-({{z}}_{\tilde{{\mathsf{a}}}}){\ms{D}}_+({{z}}_{\tilde{{\mathsf{a}}}})}{\overline{{\ms{D}}_-({{z}}_{\tilde{{\mathsf{a}}}}){\ms{N}}'_+({{z}}_{\tilde{{\mathsf{a}}}})}}
\sum\nolimits_{{\mathsf{b}}=1}^{{N^{{\mc{B}}}}}\frac{\overline{{\ms{D}}_-({{z}}_{{\mathsf{b}}}){\ms{N}}_+({{z}}_{{\mathsf{b}}})}}{{\ms{N}}'_-({{z}}_{{\mathsf{b}}}){\ms{D}}_+({{z}}_{{\mathsf{b}}})}\frac{{{z}}_{\tilde{{\mathsf{a}}}}\overline{{z}}_{{\mathsf{b}}}}{|{{z}}_{{\mathsf{b}}}-{{z}}_{\tilde{{\mathsf{a}}}}|^2}\\
&=-{\frac{1}{2}}\sum\nolimits_{\tilde{{\mathsf{b}}}=1}^{N^{{\mc{B}}}}\frac{{\ms{N}}_-({{z}}_{\tilde{{\mathsf{b}}}}){\ms{D}}_+({{z}}_{\tilde{{\mathsf{b}}}})}{\overline{{\ms{D}}_-({{z}}_{\tilde{{\mathsf{b}}}}){\ms{N}}'_+({{z}}_{\tilde{{\mathsf{b}}}})}}
\sum\nolimits_{{\mathsf{a}}=1}^{{N^{{\mc{A}}}}}\frac{\overline{{\ms{D}}_-({{z}}_{{\mathsf{a}}}){\ms{N}}_+({{z}}_{{\mathsf{a}}})}}{{\ms{N}}'_-({{z}}_{{\mathsf{a}}}){\ms{D}}_+({{z}}_{{\mathsf{a}}})}\frac{{{z}}_{\tilde{{\mathsf{b}}}}\overline{{z}}_{{\mathsf{a}}}}{|{{z}}_{{\mathsf{a}}}-{{z}}_{\tilde{{\mathsf{b}}}}|^2}.
\end{eqn}
For type II junction, the first term in \eqref{def_g_IA_symm} needs to be replaced by ${\frac{1}{2}}\sum\nolimits_\text{{odd modes}}1$.
The graphical plot of conductance entries ${\mathscr{G}}_{{\mc{A}}{\mc{I}}}$ ($={\mathscr{G}}_{{\mc{B}}{\mc{I}}}$) and ${\mathscr{G}}_{{\mc{A}}{\mc{B}}}$ are shown in Fig. \ref{latticestrip_Conductance_sq_symm} (a correspondence, and contrast due to symmetry, with (a), (b) of Fig. \ref{latticestrip_Conductance_sq_X_R_XR_smallN} and (a'), (b') of Fig. \ref{latticestrip_Conductance_sq_c_X_R_XR_smallN} is natural).

\subsection{Application to out-of-plane phonon dominated thermal transport}
{In the $3\times3$ thermal conductance matrix for the three terminals} (Fig. \ref{conductance_bifurcated_sq}), ${\sigma}_{\alpha\beta}$
 \cite{Rego1998,Angelescu1998,Blencowe1999,WangJS2008}, {where ${\alpha},{\beta}$ belong to the set $\{{\mc{I}}, {{\mc{A}}}, {{\mc{B}}}\}$}, is given by (using the physical frequency $\omega={{c}_s}{\upomega}/{\mathrm{b}}$ \cite{Bls0}, and the definition $\beta=\hbar {{c}_s}/ {\mathrm{k}_B} T {\mathrm{b}}$)
\begin{eqn}
{\sigma}_{\alpha\beta}={\sigma}_{\beta\alpha}&=\frac{\hbar}{2\pi}\int_0^\infty {\mathscr{G}}_{\beta\alpha}({\upomega})\omega\pd{f(\omega,T)}{T}d\omega
\\&
=\frac{3\beta^3{\sigma}_Q}{\pi^2}\int_0^\infty {\mathscr{G}}_{\beta\alpha}({\upomega})\frac{{\upomega}^2 exp(\beta {\upomega})}{(exp(\beta{\upomega}) -1)^2}d{\upomega},
\label{sigma_eq}
\end{eqn}
where ${\mathscr{G}}_{\beta\alpha}$ is the sum of phonon transmission probability, i.e. $\sum\nolimits_{\tilde{\alpha}=1}^{N^{\alpha}}{\ms{T}}_{\beta}(\tilde{\alpha}),$ over all allowed modes from terminal $\alpha$ to terminal $\beta$, as derived in the previous section. 
Also $f(\omega,T) = (exp(\hbar \omega/ {\mathrm{k}_B} T) -1)^{-1}$ is Bose-Einstein distribution function \cite{Rego1998,Angelescu1998} for heat carriers at the terminal. Typically ${\sigma}$ is expressed in terms of the unit quantum \cite{Rego1998,Angelescu1998} of thermal conductance ${\sigma}_Q=\pi^2{\mathrm{k}_B}^2T/(3h)$. 
The expression \eqref{sigma_eq} for the thermal conductance 
contains the sum over the 
the incident modes, say indexed by $m$, propagating in the $\alpha$ terminal and the integration over the interval $[\omega_{m;\min}, \omega_{m;\max}]$ as the pass band of the $m$th mode. Also $1/({\mathrm{k}_B} T)$, ${\mathrm{k}_B}$ is the Boltzmann constant, $T$ is the (absolute) temperature, and $\hbar$ is Planck's constant. 
The effect of scattering is captured by the presence of conductance function ${\mathscr{G}}_{\beta\alpha}$.
For perfect (ballistic) transmission, above expression \eqref{sigma_eq} equals 
\begin{eqn}
{\sigma}^{{\text{ball}}}_{\beta\alpha}&=\frac{3\beta^3{\sigma}_Q}{\pi^2}\int_0^\infty {\mathscr{G}}^{{\text{ball}}}_{\beta\alpha}({\upomega})\frac{{\upomega}^2exp(\beta {\upomega})}{(exp(\beta{\upomega}) -1)^2}d{\upomega}.
\label{sigmaballistic_eq}
\end{eqn}
Let
$${\overline{{\sigma}}_{{{\mc{A}}}{\mc{I}}}}{:=}\frac{{\sigma}_{{{\mc{A}}}{\mc{I}}}}{{\sigma}^{{\text{ball}}}_{{{\mc{A}}}{\mc{I}}}}, {\overline{{\sigma}}_{{{\mc{B}}}{\mc{I}}}}{:=}\frac{{\sigma}_{{{\mc{B}}}{\mc{I}}}}{{\sigma}^{{\text{ball}}}_{{{\mc{B}}}{\mc{I}}}}, {\overline{{\sigma}}_{{{\mc{A}}}{{\mc{B}}}}}{:=}\frac{{\sigma}_{{{\mc{A}}}{{\mc{B}}}}}{{\sigma}^{{\text{ball}}}_{{{\mc{A}}}{{\mc{B}}}}}.$$
As illustrations, Fig. \ref{latticestrip_ThermConductRelative_sq_XRXR} and Fig. \ref{latticestrip_ThermConductRelative_sq_c_XRXR} depict the temperature dependence of ${\sigma}_{{{\mc{A}}}{\mc{I}}}, {\sigma}_{{{\mc{B}}}{\mc{I}}}, {\sigma}_{{{\mc{A}}}{{\mc{B}}}}$ as well as $\overline{{\sigma}}_{{{\mc{A}}}{\mc{I}}}, \overline{{\sigma}}_{{{\mc{B}}}{\mc{I}}}, \overline{{\sigma}}_{{{\mc{A}}}{{\mc{B}}}}$.
In the case of asymmetry, whenever ${\mathtt{N}_{\mathfrak{a}}}\ne{\mathtt{N}_{\mathfrak{b}}},$ a natural application to thermal rectifiers exists. Due to availability of a succint closed form expression for the conductance function ${\mathscr{G}}_{\beta\alpha}$ presented in the previous section, the corresponding analysis is more amenable and less intensive 
in comparison to the more common computational approach \cite{Ouyang2010,xie2012nonlinear}.

\subsection{Application to electronic transport}
\label{Emodel}

Over the last two decades, electronic transport through three-terminal junction devices has received huge attention \cite{Csontos2002} owing to their non-linear electrical properties. The formalism based on the scattering-matrix method (Landauer-B{\"{u}}ttiker viewpoint 
 \cite{Landauer1957,Landauer,Buttiker1986,Buttiker1985}) takes quantum interference effects into account while incorporating 
 large flexibility in the modelling of arbitrary potential profiles. 
The waveguides with discrete Dirichlet boundary condition are natural for the electronic transport, for instance the case corresponding to Fig. \ref{latticestrip_sq_BCs_asymm_semiinfdefect}(a').
Following \cite{Thomasbook}, 
for small differences of the electrochemical potentials ${\upmu}_{\beta} -{\upmu}_{\alpha},$ in the linear response regime, the current in $\alpha$ terminal (with $f$ as the Fermi-Dirac distribution function and $g_s$ as spin factor) is
$I_{{\alpha}} = g_s \frac{|e|}{h}\sum\nolimits_{\beta, \beta\ne\alpha} \int d\upmu {\mathscr{G}}_{\alpha\beta}(\upmu) \pd{f_{\alpha}(\upmu)}{{\upmu}}({\upmu}_\beta-{\upmu}_\alpha)
=-\frac{1}{|e|}\sum\nolimits_{\beta, \beta\ne\alpha} {G}_{\alpha\beta}({\upmu}_\beta-{\upmu}_\alpha).$
Denoting the voltage differences between contacts ${\alpha}$ and ${\beta}$ as $V_{\alpha} - V_{\beta}$, 
it is seen that
$I_{{\alpha}} = -\sum\nolimits_{\beta, \beta\ne\alpha} {G}_{\alpha\beta}(V_\alpha-V_\beta),$
where the off-diagonal elements of the conductance $\Ttt{G}$ (with symmetry due to time-reversal symmetry) are 
${G}_{\alpha\beta}={G}_{\beta\alpha}=- g_s \frac{e^2}{h}\int d\upmu {\mathscr{G}}_{\alpha\beta}(\upmu) \pd{f_{\beta}(\upmu)}{{\upmu}}.$
The unit of electronic quantum conductance becomes ${\mathscr{G}}_Q=\frac{e^2}{\pi\hbar}.$ 
Thus, 
a calculation of the electrical conductance from the asymptotic form of the scattering states \cite{Nazarovbook2009} can be done, in exactly the same way as carried out earlier in this paper; eventually yielding an extremely simple expression given by the form \eqref{genGexp}. 

\section{Concluding remarks}
\label{concl}
{In this paper}, a succinct closed form expression for the transmission matrix based conductance is provided for the bifurcated discrete waveguides of square lattice.
It is found that the three-terminal Landauer--B{\"{u}}ttiker conductance across the terminals can be adjusted by modifying the number of channels and the type of confinement in the waveguides.
Overall, the presented work can be seen as a prototypical example of recently presented applications in the honeycomb \cite{Bls5c_tube,Bls5k_tube} structures as well; in fact \cite{Bls5c_tube} deals with the electronic counterpart of the problem and its results have been also reported in popular media \cite{Bls5c_tube_media}.
The discrete paradigm of the bifurcated waveguides, introduced in the paper, {is anticipated to encompass several applications in futuristic technological devices, engineering, and science that involve elastic, phononic, or electronic transport at nanoscale.}
It is anticipated to provide an analytical tool for the study of special 
Y-shaped three-terminal ballistic junction, which may play an important role in the design 
of 
nanoscale devices. 

{\bf Acknowledgments}
{\small 
\hspace{.1in}
The partial support provided by IITK/ME/20090027 and SERB MATRICS grant MTR/2017/000013 is gratefully acknowledged. 
}

\begin{appendix}
\section{Wave incident from the bifurcated portion}
\label{app_exactsol}
The discrete Fourier transform ${\mathtt{u}}^F$ of $\{{\mathtt{u}}_m\}_{m\in\Z}$ is {defined by} \begin{eqn}{\mathtt{u}}_{{\mathtt{y}}}^F={\mathtt{u}}_{{\mathtt{y}};+}+{\mathtt{u}}_{{\mathtt{y}};-}, 
\\
{\mathtt{u}}_{{\mathtt{y}};+}({{z}})=\sum\limits_{{\mathtt{x}}=0}^{+\infty} {\mathtt{u}}_{{\mathtt{x}}, {\mathtt{y}}}{{z}}^{-{\mathtt{x}}}, {\mathtt{u}}_{{\mathtt{y}};-}({{z}})=\sum\limits_{{\mathtt{x}}=-\infty}^{-1} {\mathtt{u}}_{{\mathtt{x}}, {\mathtt{y}}}{{z}}^{-{\mathtt{x}}}.\label{unpm}\end{eqn}
By \eqref{dHelmholtz_sq},
\begin{eqn}
{{\mathpzc{Q}}}({{z}}){\mathtt{u}}_{{\mathtt{y}}}^F({{z}})-({\mathtt{u}}_{{\mathtt{y}}+1}^F({{z}})+{\mathtt{u}}_{{\mathtt{y}}-1}^F({{z}}))=0, 
\\
{{\mathpzc{Q}}}({{z}}){:=}4-{{z}}-{{z}}^{-1}-{\upomega}^2, {{\lambda}}{:=}\frac{{{\mathpzc{r}}}-{{\mathpzc{h}}}}{{{\mathpzc{r}}}+{{\mathpzc{h}}}},
\\
{{\mathpzc{h}}}{:=}\sqrt{{\mathpzc{H}}}, {{\mathpzc{r}}}{:=}\sqrt{{\mathpzc{R}}}, {{\mathpzc{H}}}{:=} {\mathpzc{Q}}-2, {{\mathpzc{R}}}{:=} {\mathpzc{Q}}+2. 
\label{dHelmholtzF_sq}
\end{eqn}
The complex functions ${{\mathpzc{H}}}$, ${{\mathpzc{R}}}$, and ${\lambda}$ are defined \cite{Slepyanbook} on $\C\setminus{\ms{B}}$ where ${\ms{B}}$ denotes the union of branch cuts for ${{\lambda}}$, borne out of the chosen branch
$-\pi<\arg {{\mathpzc{H}}}({{z}}) <\pi, \Re {{\mathpzc{h}}}({{z}})>0, \Re {{\mathpzc{r}}}({{z}})>0, {\text{\rm sgn}} \Im {{\mathpzc{h}}}({{z}})={\text{\rm sgn}} \Im {{\mathpzc{r}}}({{z}}),$
for ${{\mathpzc{h}}}$ and ${{\mathpzc{r}}}$ such that 
$|{{\lambda}}({{z}})|\le1, 
 {{z}}\in\C\setminus{\ms{B}},$ as ${\upomega}_2=\Im{\upomega}$ is positive. 
The general solution of \eqref{dHelmholtzF_sq} is given by the expression
\begin{eqn}
{\mathtt{u}}_{{\mathtt{y}}}^F({{z}})=P({{z}}){{\lambda}}({{z}})^{{\mathtt{y}}}+Q({{z}}){{\lambda}}({{z}})^{-{\mathtt{y}}}, {{z}}\in{{\ms{A}}},
\label{gensol_sq}\end{eqn}
where $P, Q$ are arbitrary analytic functions on suitable annulus ${{\ms{A}}}$ around the origin in the complex plane.
The discrete Fourier transform ${\mathtt{u}}_{{\mathtt{y}}}^F$ {\eqref{unpm}} of the sequence $\{{\mathtt{u}}_{{\mathtt{x}}, {\mathtt{y}}}\}_{{\mathtt{x}}\in\Z}$ is well defined for all relevant values of ${\mathtt{y}}$ \cite{Bls2}.

A brief account of the formulation is provided {below} that indicates certain modifications in the analysis of the discrete bifurcated waveguide problem when the wave is incident from the bifurcated part of the waveguide. 
The scattering of a wave incident from any of the two bifurcated portions occurs due to the intact bonds between two fictitious separated waveguides 
above and below the bifurcation.

Due to their frequent appearance, it is useful to define 
\begin{eqn}
{{z}}_{{P}}{:=} e^{-i{\upkappa}_x}\in\C, \delta_{D+}({{z}}){:=}\sum\nolimits_{n=0}^{+\infty}{{z}}^{-n}, {{z}}\in\C, \\
\delta_{D-}({{z}}){:=}\sum\nolimits_{n=-\infty}^{-1}{{z}}^{-n}, {{z}}\in\C.
\label{zPdef_sq}
\end{eqn}

\subsection{Type I bifurcation}
\label{appsol_k}
Using the multiplicative factorization $\sLNsq_{{}}=\sLNsq_{{}+}\sLNsq_{{}-}$, the Wiener--Hopf equation (7.8) of 
\paperone{}, involving the Fourier transform of the unknown ${{\mathrm{v}}}$ (i.e., ${{\mathrm{v}}}_{{\mathtt{x}}}={{\mathrm{u}}}_{{\mathtt{x}},0}-{{\mathrm{u}}}_{{\mathtt{x}},-1}, {\mathtt{x}}\in\Z$), can be expressed as ${{\sLNsq}_{{}+}^{-1}({{z}})}{{{\mathrm{v}}}_{+}({{z}})}+{\sLNsq}_{{}-}({{z}}){{\mathrm{v}}}_-({{z}})={{\mathpzc{C}}}({{z}}), {{z}}\in{{\ms{A}}}$ where
${{\mathpzc{C}}}({{z}})=(\frac{1}{{\sLNsq}_{{}+}({{z}})}-{\sLNsq}_{{}-}({{z}})){{\mathrm{A}}}{{{\mathrm{v}}}_{({{\kappa}^{i}})}}(-1)\delta_{D+}({{z}} {{z}}_{{P}}^{-1}), 
 {{z}}\in{{\ms{A}}},$ 
and ${\sLNsq_{}}={\ms{N}}/{\ms{D}}$ with outer boundary condition dependent ${\ms{N}}$ and ${\ms{D}}$ as given in Table \ref{bifurcatedstripbc_sq_ND}. For short form of expression, ${{{\mathrm{v}}}_{({{\kappa}^{i}})}}{:=}({\seigen}_{({{\kappa}^{i}}){{{\mathtt{N}_{\mathfrak{b}}}+1}}}-{\seigen}_{({{\kappa}^{i}}){{{\mathtt{N}_{\mathfrak{b}}}}}})$.
An additive factorization \cite{Noble} ${{\mathpzc{C}}}={{\mathpzc{C}}}_{+}({{z}})+{{\mathpzc{C}}}_{-}({{z}})$ with
\begin{eqn}
{{\mathpzc{C}}}_\pm({{z}})&=
\pm{{\mathrm{A}}}{{{\mathrm{v}}}_{({{\kappa}^{i}})}}({\sLNsq}_{{}-}({{z}}_{{P}})-{\sLNsq}_{{}\pm}^{\mp1}({{z}}))
\\&
\delta_{D+}({{z}} {{z}}_{{P}}^{-1})\delta_{{s},{B}}, {{z}}\in{\ms{A}},
\label{CpmK_sq}
\end{eqn}
and a reasoning based on the Liouville's theorem, indeed, entirely identical to that applied for incidence ahead of the bifurcation \cite{Bls9s}, leads to the solution of the discrete Wiener--Hopf equation, 
in terms of the one-sided discrete Fourier transform, as
\begin{eqn}
{{\mathtt{v}}}_{\pm}({{z}})&={{\mathpzc{C}}}_{\pm}({{z}})\sLNsq_{\pm}^{\pm1}({{z}}),
\\&
{{z}}\in\C, |{{z}}|\gtrless
\bfrac{\max}{\min}
\{{\radius}_{\pm}, {\radius}_{L_{{}}}^{\pm1}\}.
\label{vpm_sq}
\end{eqn}

The complex function ${\mathtt{v}}^F$ as a complete solution for both directions of incidence (indicated by ${\mathfrak{s}}={A}$ for incidence from portion ahead of bifurcation and ${\mathfrak{s}}={B}$ for either of the portions in the bifurcated region) is given by
\begin{eqn}
{{\mathrm{v}}}^F({{z}})&={{\mathrm{A}}}{\mathtt{C}}_0\frac{{{z}} {{\mathpzc{K}}}({{z}})}{{{z}}-{{z}}_{{P}}}, 
{{\mathpzc{K}}}({{z}}){:=}\frac{1-{\sLNsq}_{{}}({{z}})}{{\sLNsq}_{{}-}({{z}})}, \text{where }
\\&
{\mathtt{C}}_0{:=}
-\frac{{{\mathrm{v}}}_{({{\kappa}^{i}})}}{{\sLNsq}_{+}({{z}}_{{P}})}\delta_{{{\mathfrak{s}}},{A}}
-{{{\mathrm{v}}}_{({{\kappa}^{i}})}}{\sLNsq}_{{}-}({{z}}_{{P}})\delta_{{{\mathfrak{s}}},{B}}
\in\C,
\label{vzsol}
\end{eqn}
for ${{z}}\in{{\ms{A}}}$. 
Since
${\mathtt{v}}_{{\mathtt{x}}}=\frac{1}{2\pi i}\oint_{\mathbb{T}} {\mathtt{v}}^F({{z}}){{z}}^{{\mathtt{x}}-1}d{{z}}, {\mathtt{x}}\in\Z, $
where ${\mathtt{v}}^F={\mathtt{v}}_++{\mathtt{v}}_-$, as ${\mathtt{x}}\to\pm\infty$, the asymptotic expression for ${\mathtt{v}}$ can be obtained by analyzing \eqref{vpm_sq} with
${\mathtt{v}}_{{\mathtt{x}}}=\frac{1}{2\pi i}\oint_{\mathbb{T}} {\mathtt{v}}_\pm({{z}}){{z}}^{{\mathtt{x}}-1}d{{z}}, {\mathtt{x}}\in\Z^{\pm}.$
Indeed, for ${\mathfrak{s}}={B}$, after deforming the contour of integration and applying residue calculus, 
the exact expression is given by 
(assuming ${\upomega}_2>0$)
\begin{eqn}
{\mathtt{v}}_{{\mathtt{x}}}=\pm\sum\limits_{{{z}}={{z}}_{{\ast}}, |{{z}}_{{\ast}}|\lessgtr1} \text{Res }{{\mathpzc{C}}}_{\pm}({{z}}){\sLNsq}_{{}\pm}^{\pm1}({{z}}){{z}}^{{\mathtt{x}}-1}, {\mathtt{x}}\in\Z^{\pm},
\label{vm0asym1_sq}
\end{eqn}
where ${{\mathpzc{C}}}_{\pm}$ is given by \eqref{CpmK_sq} and where the sets of ${z}$ corresponding to outgoing waves are given in \eqref{Zer_sq}. 
Hence,
\begin{eqn}
{\mathtt{v}}_{{\mathtt{x}}}&\sim{{\mathrm{A}}}{{{\mathrm{v}}}_{({{\kappa}^{i}})}}\frac{{\ms{N}}_-({{z}}_{{P}})}{{\ms{D}}_-({{z}}_{{P}})}\sum\limits_{{{\umode}}^+_{{\mc{I}}}
}\frac{{\ms{N}}_+({{z}})}{{\ms{D}}'_+({{z}})}\frac{{{z}}^{{\mathtt{x}}}}{{{z}}-{{z}}_{{P}}}, \\
{\mathtt{v}}_{{\mathtt{x}}}&\sim{{\mathrm{A}}}{{{\mathrm{v}}}_{({{\kappa}^{i}})}}(-{{z}}_{{P}}^{{\mathtt{x}}}+\frac{{\ms{N}}_-({{z}}_{{P}})}{{\ms{D}}_-({{z}}_{{P}})}\sum\limits_{
{{\umode}}^{-}_{{\mc{A}}{\mc{B}}}}\frac{{\ms{D}}_-({{z}})}{{\ms{N}}'_-({{z}})}\frac{{{z}}^{{\mathtt{x}}}}{{{z}}-{{z}}_{{P}}}),
\label{vm0asym}
\end{eqn}
as ${\mathtt{x}}\to+\infty, {\mathtt{x}}\to-\infty,$ respectively.
The far-field can also be determined {using the normal mode expansion as above,} 
and after solving for the coefficients, it is found that total displacement field is given by \eqref{farfield_k_sq_altinc} as ${\mathtt{x}}\to+\infty$ and ${\mathtt{x}}\to-\infty$, respectively. 

\subsection{Type II bifurcation}
\label{appsol_c}
The incident wave mode is assumed that satisfies the equation of motion of the 
appropriate bifurcated portion. 
Hence, analogous to \eqref{dHelmholtz_sq},
${\mathtt{u}}^{{t}}_{{{\mathtt{x}}}+1, {0}}+{\mathtt{u}}^{{t}}_{{{\mathtt{x}}}-1, {0}}+{\mathtt{u}}^{{t}}_{{{\mathtt{x}}}, 1}+{\mathtt{u}}^{{t}}_{{{\mathtt{x}}}, -1}+({\upomega}^2-4){\mathtt{u}}^{{t}}_{{{\mathtt{x}}}, {0}}=0, {\mathtt{x}}\ge0,$
$({\upomega}^2-4){\mathtt{u}}^{{t}}_{{{\mathtt{x}}}, {0}}=0, {\mathtt{x}}<0,$
i.e.
\begin{eqn}
{\mathtt{u}}_{{{\mathtt{x}}}+1, {0}}+{\mathtt{u}}_{{{\mathtt{x}}}-1, {0}}+{\mathtt{u}}_{{{\mathtt{x}}}, 1}+{\mathtt{u}}_{{{\mathtt{x}}}, -1}
\\
+{\mathtt{u}}^{i}_{{{\mathtt{x}}}, 1}+{\mathtt{u}}^{i}_{{{\mathtt{x}}}, -1}+({\upomega}^2-4){\mathtt{u}}_{{{\mathtt{x}}}, {0}}=0, {\mathtt{x}}\ge0,\\
({\upomega}^2-4){\mathtt{u}}_{{{\mathtt{x}}}, {0}}=({\upomega}^2-4){\mathtt{u}}^{i}_{{{\mathtt{x}}}, {0}}=0, {\mathtt{x}}<0.
\end{eqn}
After application of the discrete Fourier transform \eqref{unpm} (note ${\mathtt{u}}_{0; -}=0$), 
\begin{eqn}
-{\mathtt{u}}_{0, {0}}{z}+{\mathtt{u}}_{-1, {0}}+({\mathtt{u}}_{1;+}+{\mathtt{u}}_{-1;+})
\\
+({\mathtt{u}}^{i}_{1;+}+{\mathtt{u}}^{i}_{-1;+})=(4-({z}+{z}^{-1})-{\upomega}^2){\mathtt{u}}_{0;+},\\
({\upomega}^2-4){\mathtt{u}}_{0;-}=({\upomega}^2-4){\mathtt{u}}^{i}_{0;-}=0.
\label{sliteq_sq}
\end{eqn}
It is natural to define ${{\mathrm{w}}}^F$ and ${\mathpzc{V}}$ by
\begin{eqn}
{{\mathrm{w}}}^F({{z}})={\mathtt{u}}_{+1}^{F}({{z}})+{\mathtt{u}}_{-1}^{F}({{z}}), {{\mathrm{w}}}^F={\mathtt{u}}^F_0{\mathpzc{V}}. 
\label{V_sq}
\end{eqn}
By {a rearrangement of} 
\eqref{sliteq_sq}, {using} \eqref{V_sq} and \eqref{unpm}, it follows that ${\mathrm{w}}$ satisfies
(as the counterpart of Eq. (3.2) of 
\paperone{})
\begin{eqn}
{{\mathpzc{Q}}}{\mathtt{u}}_{0;+}&={{\mathrm{w}}}^{i}_{+}+{{\mathpzc{W}}}+{{\mathrm{w}}}_{+} \text{ on }{\ms{A}}_{{c}},
\label{sliteq_sq_altinc}
\end{eqn}
where ${\mathpzc{W}}$ is given by 
${{\mathpzc{W}}}({{z}}){:=}{\mathtt{u}}_{-1, 0}-{{z}} {\mathtt{u}}_{0, 0}=-{\mathtt{u}}^{i}_{-1, 0}-{{z}} {\mathtt{u}}_{0, 0}=-{{z}} {\mathtt{u}}_{0, 0}, {{z}}\in\C$ (i.e. Eq. (3.2c) of 
\paperone{}) as ${\mathtt{u}}^{i}_{-1, 0}=0$.
Further, upon substitution of the equation \eqref{sliteq_sq_altinc}, 
the discrete Wiener--Hopf equation for ${{\mathrm{w}}}_{+}$ and ${{\mathrm{w}}}_{-}$ is found to be (compare with Eq. (3.4a) of 
\paperone{})
\begin{eqn}
{\sLNsq_{}}{{\mathrm{w}}}_{+}+{{\mathrm{w}}}_{-}=({{\mathpzc{W}}}+{{\mathrm{w}}}^{i}_{+}){{\mathpzc{Q}}}^{-1}{\mathpzc{V}} \text{ on }{\ms{A}}_{},
\label{WHCeq_sq_gen_altinc}
\end{eqn}
with ${\sLNsq_{}}={\ms{N}}/{\ms{D}}={\ms{N}}/({\mathpzc{Q}}\mathring{{\ms{D}}})$ with outer boundary condition dependent ${\ms{N}}$ and ${\ms{D}}$ as given in Table \ref{bifurcatedstripbc_sq_ND_c}. 
Similar to \S3.5 of 
\paperone{}, further analysis can be carried out for the incidence from the bifurcated portion of the waveguides. In this case, in place of Eq. (3.15) of 
\paperone{}, with ${\mathpzc{C}}=({{\mathpzc{W}}}+{{\mathrm{w}}}^{i}_{+})({\sLNsq}_{-}^{-1}-{\sLNsq}_{+}),$
which leads to 
\begin{eqn}
{{\mathpzc{C}}}_{\pm}({{z}})
&=\mp{{z}}(-{\mathtt{u}}_{0, 0})({\sLNsq_{}}_\pm^{\pm1}({{z}})-{l_{}}_{+0})
\\&
\mp(-{\mathtt{u}}^{i}_{-1, 0})
({\sLNsq_{}}_\pm^{\pm1}({{z}})-{\overline{l}_{}}_{-0})\\
&\mp{{\mathrm{A}}}({\seigen}_{({{\kappa}^{i}}){{\mathtt{N}_{\mathfrak{b}}}+2}}+{\seigen}_{({{\kappa}^{i}}){{\mathtt{N}_{\mathfrak{b}}}}})\delta_{D+}({{z}} {{z}}_{{P}}^{-1})
\\&
\big({\sLNsq}_{\pm}^{\pm1}({{z}})-{\sLNsq}^{-1}_{-}({{z}}_{{P}})\big),
\label{Cpm_sq_gen_altinc}
\end{eqn}
where ${\overline{l}_{}}_{-0}=\lim_{{{z}}\to0}{{\sLNsq}_{}}_-^{-1}({{z}})$. Finally, the solution of \eqref{WHCeq_sq_gen_altinc} is written (same as Eq. (3.16) of 
\paperone{}) as $${{\mathrm{w}}}_{\pm}({{z}})= {{\mathpzc{C}}}_\pm({{z}}){{\sLNsq}_{}}_{\pm}({{z}})^{\mp1}, {{z}}\in\C, |{{z}}|>\bfrac{\max}{\min}\{{\radius}_\pm, {\radius}_{L_{}}^{\pm1}\}.$$ 
Using the inverse discrete Fourier transform, 
${\mathrm{w}}_{{\mathtt{x}}}=\frac{1}{2\pi i}\oint_{\mathbb{T}} {\mathrm{w}}_\pm({{z}}){{z}}^{{\mathtt{x}}-1}d{{z}}, {\mathtt{x}}\in\Z^{\pm},$
and upon substitution
of \eqref{Cpm_sq_gen_altinc}, the exact expression can be constructed. 
Thereby, the expression for ${\mathtt{u}}_{0;+}$ can be found using \eqref{sliteq_sq_altinc} 
and ${\mathtt{u}}^{{t}}_{0, 0}$ is obtained as 
${\mathtt{u}}^{{t}}_{0, 0}=-{{\mathrm{A}}}{\seigen}_{({{\kappa}^{i}}){{\mathtt{N}_{\mathfrak{b}}}+1}}{{\sLNsq}_{}}_+({{z}}_{{P}}){{\mathpzc{Q}}}({{z}}_{{P}})/({l_{}}_{+0}({{z}}_{{P}}-{{z}}_{{\mathpzc{q}}}^{-1})).$ 
Using the inverse discrete Fourier transform, 
\begin{eqn}
{\mathtt{u}}_{0,0}&=\frac{1}{2\pi i}\oint_{\mathbb{T}} {\mathtt{u}}_{{0}; +}({{z}}){{z}}^{-1}d{{z}}\\
&=\frac{1}{2\pi i}\oint_{\mathbb{T}} {{\mathpzc{Q}}}^{-1}({{z}})({{\mathrm{w}}}^{i}_{+}+{{\mathpzc{W}}}+{{\mathrm{w}}}_{+}){{z}}^{-1}d{{z}}.
\end{eqn}
Recall that
$${{\mathpzc{Q}}}({{z}})={{z}}_{{\mathpzc{q}}}^{-1}(1-{{z}}_{{\mathpzc{q}}}{{z}})(1-{{z}}_{{\mathpzc{q}}}{{z}}^{-1})=-{{z}}^{-1}({z}-{{z}}_{{\mathpzc{q}}})({{z}}-{z}_{{\mathpzc{q}}}^{-1}),$$
So
\begin{eqn}
{\mathtt{u}}_{0,0}&=\frac{1}{2\pi i}\oint_{\mathbb{T}} \frac{({{\mathrm{w}}}^{i}_{+}+{{\mathpzc{W}}}+{{\mathrm{w}}}_{+})}{-{{z}}^{-1}({z}-{{z}}_{{\mathpzc{q}}})({{z}}-{z}_{{\mathpzc{q}}}^{-1})}{{z}}^{-1}d{{z}}\\
&=\frac{1}{2\pi i}\oint_{\mathbb{T}} \frac{({{\mathrm{w}}}^{i}_{+}+{{\mathpzc{W}}}+{{\mathrm{w}}}_{+})}{-({z}-{{z}}_{{\mathpzc{q}}})({{z}}-{z}_{{\mathpzc{q}}}^{-1})}d{{z}}\\
&=(-{\mathtt{u}}^{i}_{-1, 0}-{{z}}_{{\mathpzc{q}}}^{-1} {\mathtt{u}}_{0, 0}+{{\mathrm{w}}}_{+}({{z}}_{{\mathpzc{q}}}^{-1})+{{\mathrm{w}}}^{i}_{+}({{z}}_{{\mathpzc{q}}}^{-1}))/({z}^{-1}_{{\mathpzc{q}}}-{{z}}_{{\mathpzc{q}}})\\
&=({\mathtt{u}}_{-1, 0}+{{z}}_{{\mathpzc{q}}} {\mathtt{u}}^{i}_{0, 0}-{{\mathrm{w}}}_{-}({{z}}_{{\mathpzc{q}}}))/({z}^{-1}_{{\mathpzc{q}}}-{{z}}_{{\mathpzc{q}}}).
\end{eqn}
Here
\begin{eqn}
{{\mathpzc{C}}}_{+}({{z}})&=-{{z}}(-{\mathtt{u}}_{0, 0})({\sLNsq_{}}_+({{z}})-{l_{}}_{+0})\\
&-{{\mathrm{A}}}({\seigen}_{({{\kappa}^{i}}){{\mathtt{N}_{\mathfrak{b}}}+2}}+{\seigen}_{({{\kappa}^{i}}){{\mathtt{N}_{\mathfrak{b}}}}})\delta_{D+}({{z}} {{z}}_{{P}}^{-1})
\\&
\big({\sLNsq}_{;+}({{z}})-{\sLNsq}^{-1}_{-}({{z}}_{{P}})\big),
\end{eqn}
so that
\begin{eqn}
{{\mathrm{w}}}_{+}({{z}}_{{\mathpzc{q}}}^{-1})&={{\mathpzc{C}}}_+({{z}}_{{\mathpzc{q}}}^{-1}){{\sLNsq}}^{-1}_{+}({{z}}_{{\mathpzc{q}}}^{-1})\\
&=-{{z}}_{{\mathpzc{q}}}^{-1}(-{\mathtt{u}}_{0, 0})(1-{{\sLNsq}}^{-1}_{+}({{z}}_{\mathpzc{q}}^{-1}){l_{}}_{+0})\\
&-{{\mathrm{A}}}({\seigen}_{({{\kappa}^{i}}){{\mathtt{N}_{\mathfrak{b}}}+2}}+{\seigen}_{({{\kappa}^{i}}){{\mathtt{N}_{\mathfrak{b}}}}})\delta_{D+}({{z}}_{\mathpzc{q}}^{-1} {{z}}_{{P}}^{-1})
\\&
\big(1-{{\sLNsq}}^{-1}_{+}({{z}}_{\mathpzc{q}}^{-1}){\sLNsq}^{-1}_{-}({{z}}_{{P}})\big).
\end{eqn}
Hence,
\begin{eqn}
0&=-{{z}}_{\mathpzc{q}}^{-1}{\mathtt{u}}_{0, 0}{{\sLNsq}}^{-1}_{+}({{z}}_{\mathpzc{q}}^{-1}){l_{}}_{+0}\\
&+{{\mathrm{A}}}({\seigen}_{({{\kappa}^{i}}){{\mathtt{N}_{\mathfrak{b}}}+2}}+{\seigen}_{({{\kappa}^{i}}){{\mathtt{N}_{\mathfrak{b}}}}})\delta_{D+}({{z}}_{\mathpzc{q}}^{-1} {{z}}_{{P}}^{-1})
\\&
{{\sLNsq}}^{-1}_{+}({{z}}_{\mathpzc{q}}^{-1}){\sLNsq}^{-1}_{-}({{z}}_{{P}}),
\end{eqn}
which gives
${\mathtt{u}}_{0, 0}={{\mathrm{A}}}\frac{{\seigen}_{({{\kappa}^{i}}){{\mathtt{N}_{\mathfrak{b}}}+2}}+{\seigen}_{({{\kappa}^{i}}){{\mathtt{N}_{\mathfrak{b}}}}}}{{l_{}}_{+0}({{z}}_{\mathpzc{q}}^{-1}-{{z}}_{{P}}){\sLNsq}_{-}({{z}}_{{P}})}.$
The complex function ${\mathtt{u}}_{0}^F$ as a complete solution on the bifurcated rowfor both directions of incidence (indicated by ${\mathfrak{s}}={A}$ for incidence from portion ahead of bifurcation and ${\mathfrak{s}}={B}$ for either of the portions in the bifurcated region) is given by
\begin{eqn}
{\mathtt{u}}_{0}^F({{z}})&={{\mathrm{A}}}{\mathtt{C}}_0\frac{{{z}}{{\mathpzc{K}}}({{z}})}{{{z}}-{{z}}_{{P}}}, 
{{\mathpzc{K}}}({{z}}){:=}\frac{1}{(1-{{z}}_{\mathpzc{q}}{{z}}^{-1}){{\sLNsq}_{}}_+({{z}})}, {{z}}\in{{\ms{A}}},
\\&
\text{where }{\mathtt{C}}_0{:=}{\seigen}_{({{\kappa}^{i}}){{\mathtt{N}_{\mathfrak{b}}}+1}}{{{\mathpzc{Q}}}({{z}}_{{P}}){{\sLNsq}_{}}_+({{z}}_{{P}})}({{z}}_{\mathpzc{q}}^{-1}-{{z}}_{{P}})^{-1}\delta_{{{\mathfrak{s}}},{A}}\\&
{+\frac{{\seigen}_{({{\kappa}^{i}}){{\mathtt{N}_{\mathfrak{b}}}+2}}+{\seigen}_{({{\kappa}^{i}}){\mathtt{N}_{\mathfrak{b}}}}}{({{z}}_{{\mathpzc{q}}}^{-1}-{z}_{P}){\sLNsq}_{-}({z}_{P})}\delta_{{{{\mathfrak{s}}}},{B}}}\in\C.
\label{uzsol}
\end{eqn}
{Notice that irrespective of the direction of incidence, ${\mathtt{C}}_0={l}_{+0}{{\mathrm{A}}}^{-1}{{\mathtt{u}}}^{t}_{0, 0}.$}

Further, for ${\mathfrak{s}}={B}$, a far-field approximation yields 
\begin{eqn}
{\mathrm{w}}_{{\mathtt{x}}}
&\sim{{\mathrm{A}}}({\seigen}_{({{\kappa}^{i}}){{\mathtt{N}_{\mathfrak{b}}}+2}}+{\seigen}_{({{\kappa}^{i}}){{\mathtt{N}_{\mathfrak{b}}}}})\frac{{\ms{N}}_+({{z}}_{{P}})}{{\ms{D}}_+({{z}}_{{P}})}
\frac{1}{{{z}}_{{P}}-{{z}}_{\mathpzc{q}}^{-1}}
\\&
\sum\nolimits_{{{z}}\in{\umode}^+_{{\mc{I}}}}\frac{{{z}}-{{z}}_{\mathpzc{q}}^{-1}}{{{z}}-{{z}}_{{P}}}\frac{{\ms{D}}_+({{z}})}{{\ms{N}}'_+({{z}})}{{z}}^{{\mathtt{x}}}, \\
{\mathrm{w}}_{{\mathtt{x}}}&\sim{{\mathrm{A}}}({\seigen}_{({{\kappa}^{i}}){{\mathtt{N}_{\mathfrak{b}}}+2}}+{\seigen}_{({{\kappa}^{i}}){{\mathtt{N}_{\mathfrak{b}}}}})\frac{{\ms{N}}_+({{z}}_{{P}})}{{\ms{D}}_+({{z}}_{{P}})}
\frac{1}{{{z}}_{{P}}-{{z}}_{\mathpzc{q}}^{-1}}
\\&
\sum\nolimits_{{{z}}\in{\umode}^-_{{\mc{A}}{\mc{B}}}}\frac{{{z}}-{{z}}_{\mathpzc{q}}^{-1}}{{{z}}-{{z}}_{{P}}}\frac{{\ms{N}}_-({{z}})}{{\ms{D}}'_-({{z}})}{{z}}^{{\mathtt{x}}}\\
&-{{\mathrm{A}}}({\seigen}_{({{\kappa}^{i}}){{\mathtt{N}_{\mathfrak{b}}}+2}}+{\seigen}_{({{\kappa}^{i}}){{\mathtt{N}_{\mathfrak{b}}}}})(1-{{\sLNsq}_{}}({{z}}_{{P}})){{z}}_{{P}}^{{\mathtt{x}}}, 
\label{wsolpm}
\end{eqn}
as ${\mathtt{x}}\to+\infty, {\mathtt{x}}\to-\infty,$ respectively, where the sets of ${z}$ corresponding to outgoing waves are given in \eqref{Zer_sq}.
The far-field can also be determined {using the normal mode expansion as above,} 
and after solving for the coefficients, it is found that total displacement field is given by \eqref{farfield_c_sq}
as ${\mathtt{x}}\to+\infty$ and ${\mathtt{x}}\to-\infty$, respectively. 

\section{Scattering matrix for three terminal junction}
\label{app_scatter}
The
far-field in a bifurcated waveguide (see Fig. \ref{latticestrip_sq_BCs_asymm_semiinfdefect}) can be determined (suitably) in terms of the normal modes associated with the different portions (ahead, indicated by subscript $({{\mathsf{i}}})$, and behind, by $({{\mathsf{a}}})$ and $({{\mathsf{b}}})$) of the bifurcated lattice waveguides (see Fig. \ref{latticestrip_sq_BCs_asymm_semiinfdefect}). The normal modes\footnote{See Footnote \ref{note_npp}.} for a square lattice waveguide with fixed or free boundary are well known (see \cite{Bls9} for an elaborate list).
Recall that \cite{Bls9s} for the {\em incidence from the portion ahead} (see Fig. \ref{conductance_bifurcated_sq} top) of the bifurcation ({Ignore the superscript ${t}$ on ${\mathtt{u}}$ in contrast to the notation for the total wave field earlier.}), the transmitted wave ahead and behind the bifurcation tip, respectively, is given by, as ${\mathtt{x}}\to+\infty$ and ${\mathtt{x}}\to\infty,$ respectively,
\begin{eqn}
{\mathtt{u}}_{{\mathtt{x}}, {\mathtt{y}}}^{{\mc{I}}}\sim{{\mathrm{A}}}{\seigen}_{({\tilde{{{\mathsf{i}}}}}){{\nu}}}{{z}}_{{\tilde{{{\mathsf{i}}}}}}^{{\mathtt{x}}}+{{\mathrm{A}}}\sum\nolimits_{{{\mathsf{i}}}=1}^{{N^{{\mc{I}}}}}{{\ms{T}}}_{{\tilde{{{\mathsf{i}}}}}{{\mathsf{i}}}}{\seigen}_{({{\mathsf{i}}}){{\nu}}}{{z}}_{{{\mathsf{i}}}}^{{\mathtt{x}}}, \\
\text{ and }
{\mathtt{u}}_{{\mathtt{x}}, {\mathtt{y}}}^{{\mc{A}}}\sim{{\mathrm{A}}}\sum\nolimits_{{{\mathsf{a}}}=1}^{{N^{{\mc{A}}}}}{{\ms{T}}}_{{\tilde{{{\mathsf{i}}}}}{{\mathsf{a}}}}{\seigen}_{({{\mathsf{a}}}){{\nu}}}{{z}}_{{\mathsf{a}}}^{{\mathtt{x}}}, \\
{\mathtt{u}}_{{\mathtt{x}}, {\mathtt{y}}}^{{\mc{B}}}\sim{{\mathrm{A}}}\sum\nolimits_{{{\mathsf{b}}}=1}^{{N^{{\mc{B}}}}}{{\ms{T}}}_{{\tilde{{{\mathsf{i}}}}}{{\mathsf{b}}}}{\seigen}_{({{\mathsf{b}}}){{\nu}}}{{z}}_{{\mathsf{b}}}^{{\mathtt{x}}}. \label{assumedfarfield_sq}
\end{eqn}
It is prudent to define the coefficients in above expression, so called reflection and transmission amplitudes as appropriate \cite{Nazarovbook2009}, in the following manner:
\begin{eqn}
{{\tau}}^{{\mc{I}}\tilde{{\mc{I}}}}_{{{\mathsf{i}}}{\tilde{{{\mathsf{i}}}}}}={{\ms{T}}}_{{\tilde{{{\mathsf{i}}}}}{{\mathsf{i}}}}\sqrt{\frac{|{\velG}_{{{\mathsf{i}}}}|}{|{\velG}_{{\tilde{{{\mathsf{i}}}}}}|}},
{{\tau}}^{{{\mc{A}},{\mc{B}}}\tilde{{\mc{I}}}}_{{{{\mathsf{a}}},{{\mathsf{b}}}}{\tilde{{{\mathsf{i}}}}}}={{\ms{T}}}_{{\tilde{{{\mathsf{i}}}}}{{{\mathsf{a}}},{{\mathsf{b}}}}}\sqrt{\frac{|{\velG}_{{{{\mathsf{a}}},{{\mathsf{b}}}}}|}{|{\velG}_{{\tilde{{{\mathsf{i}}}}}}|}},
\label{scaamp_sq}
\end{eqn}
where ${\velG}$ denotes the group velocity of the indicated (by subscript) wave mode.
Using the subscript on ${\mathtt{u}}$ to denote the {incident wave channel} and the superscript to denote the portion into which the wave field is asymptotically evaluated, 
\eqref{assumedfarfield_sq} can be expressed in this notation, as ${\mathtt{x}}\to+\infty$ and ${\mathtt{x}}\to\infty,$ respectively, in the following manner
\begin{eqn}
{\mathtt{u}}^{{\mc{I}}}_{\tilde{{{\mathsf{i}}}}}\sim{\mathtt{u}}_{{\tilde{{{\mathsf{i}}}}}}+\sum\nolimits_{{{{\mathsf{i}}}}=1}^{N^{{\mc{I}}}}{{\tau}}_{{{\mathsf{i}}}{{\tilde{{{\mathsf{i}}}}}}}\sqrt{\frac{|{\velG}_{{\tilde{{{\mathsf{i}}}}}}|}{|{\velG}_{{{\mathsf{i}}}}|}}{\mathtt{u}}_{{{\mathsf{i}}}}, 
\\
\text{ and }
{\mathtt{u}}^{{{\mc{A}},{\mc{B}}}}_{\tilde{{{\mathsf{i}}}}}\sim\sum\nolimits_{{{{{\mathsf{a}}},{{\mathsf{b}}}}}=1}^{N^{{{\mc{A}},{\mc{B}}}}}{{\tau}}_{{{{\mathsf{a}}},{{\mathsf{b}}}}{\tilde{{{\mathsf{i}}}}}}\sqrt{\frac{|{\velG}_{\tilde{{{\mathsf{i}}}}}|}{|{\velG}_{{{{\mathsf{a}}},{{\mathsf{b}}}}}|}}{\mathtt{u}}_{{{{\mathsf{a}}},{{\mathsf{b}}}}}. 
\label{newfarfield_sq}
\end{eqn}

Also for the {\em incidence from the bifurcated portion} (see Fig. \ref{conductance_bifurcated_sq} center and bottom corresponding to $\tilde{{\mathsf{a}}}$ and $\tilde{{\mathsf{b}}}$, respectively), 
the transmitted wave ahead and behind the bifurcation tip, respectively, is given by
\begin{eqn}
{\mathtt{u}}^{{\mc{I}}}_{{\mathtt{x}}, {\mathtt{y}}}\sim{{\mathrm{A}}}\sum\nolimits_{{{\mathsf{i}}}=1}^{{N^{{\mc{I}}}}}{{\ms{T}}}_{{\tilde{{\mathsf{a}}},\tilde{{\mathsf{b}}}}{{\mathsf{i}}}}{\seigen}_{({{\mathsf{i}}}){{\nu}}}{{z}}_{{{\mathsf{i}}}}^{{\mathtt{x}}}, {\mathtt{x}}\to+\infty, \\
{\mathtt{u}}_{{\mathtt{x}}, {\mathtt{y}}}^{{\mc{A}},{\mc{B}}}\sim{{\mathrm{A}}}{\seigen}_{({\tilde{{\mathsf{a}}},\tilde{{\mathsf{b}}}}){{\nu}}}{{z}}_{{\tilde{{\mathsf{a}}},\tilde{{\mathsf{b}}}}}^{{\mathtt{x}}}+{{\mathrm{A}}}\sum\nolimits_{{{\mathsf{a}}}=1}^{{N^{{\mc{A}}}}}{{\ms{T}}}_{{\tilde{{\mathsf{a}}},\tilde{{\mathsf{b}}}}{{{\mathsf{a}}},{{\mathsf{b}}}}}{\seigen}_{({{{\mathsf{a}}},{{\mathsf{b}}}}){{\nu}}}{{z}}_{{{\mathsf{a}}},{{\mathsf{b}}}}^{{\mathtt{x}}},\\
{\mathtt{u}}_{{\mathtt{x}}, {\mathtt{y}}}^{{\mc{B}},{\mc{A}}}\sim{{\mathrm{A}}}\sum\nolimits_{{{\mathsf{b}}}=1}^{{N^{{\mc{B}}}}}{{\ms{T}}}_{{\tilde{{\mathsf{a}}},\tilde{{\mathsf{b}}}}{{\mathsf{b}}},{{\mathsf{a}}}}{\seigen}_{({{\mathsf{b}}},{{\mathsf{a}}}){{\nu}}}{{z}}_{{{\mathsf{b}}},{{\mathsf{a}}}}^{{\mathtt{x}}}, {\mathtt{x}}\to-\infty.
\label{assumedfarfield_sq_altinc}
\end{eqn}
{Using a concise notation, above can be expressed as}
\begin{eqn}
{\mathtt{u}}^{{{\mc{A}},{\mc{B}}}}_{\tilde{{\mathsf{a}}},\tilde{{\mathsf{b}}}}={\mathtt{u}}_{\tilde{{\mathsf{a}}},\tilde{{\mathsf{b}}}}+\sum\nolimits_{{{{\mathsf{a}}},{{\mathsf{b}}}}=1}^{N^{{{\mc{A}},{\mc{B}}}}}{{\tau}}_{{{{\mathsf{a}}},{{\mathsf{b}}}}{\tilde{{\mathsf{a}}},\tilde{{\mathsf{b}}}}}\sqrt{\frac{|{\velG}_{{\tilde{{\mathsf{a}}},\tilde{{\mathsf{b}}}}}|}{|{\velG}_{{{{\mathsf{a}}},{{\mathsf{b}}}}}|}}{\mathtt{u}}_{{{{\mathsf{a}}},{{\mathsf{b}}}}}, 
\\
{\mathtt{u}}^{{\mc{B}},{\mc{A}}}_{\tilde{{\mathsf{a}}},\tilde{{\mathsf{b}}}}=\sum\nolimits_{{{\mathsf{b}},{\mathsf{a}}}=1}^{N^{{\mc{B}},{\mc{A}}}}{{\tau}}_{{{\mathsf{b}},{\mathsf{a}}}{\tilde{{\mathsf{a}}},\tilde{{\mathsf{b}}}}}\sqrt{\frac{|{\velG}_{{\tilde{{\mathsf{a}}},\tilde{{\mathsf{b}}}}}|}{|{\velG}_{{{\mathsf{b}},{\mathsf{a}}}}|}}{\mathtt{u}}_{{{\mathsf{b}},{\mathsf{a}}}}, 
\\
{\mathtt{u}}^{{\mc{I}}}_{\tilde{{\mathsf{a}}},\tilde{{\mathsf{b}}}}=\sum\nolimits_{{{\mathsf{i}}}=1}^{N^{{\mc{I}}}}{{\tau}}_{{{\mathsf{i}}}{\tilde{{\mathsf{a}}},\tilde{{\mathsf{b}}}}}\sqrt{\frac{|{\velG}_{\tilde{{\mathsf{a}}},\tilde{{\mathsf{b}}}}|}{|{\velG}_{{{\mathsf{i}}}}|}}{\mathtt{u}}_{{{\mathsf{i}}}}. 
\label{newfarfield_sq_altinc}
\end{eqn}
Similar to \eqref{scaamp_sq}, it is natural to {consider the relations}
\begin{eqn}
{{\tau}}^{{\mc{I}}{\tilde{{\mc{A}}},\tilde{{\mc{B}}}}}_{{{\mathsf{i}}}{\tilde{{\mathsf{a}}},\tilde{{\mathsf{b}}}}}={{\ms{T}}}_{{\tilde{{\mathsf{a}}},\tilde{{\mathsf{b}}}}{{\mathsf{i}}}}\sqrt{\frac{|{\velG}_{{{{{\mathsf{i}}}}}}|}{|{\velG}_{\tilde{{\mathsf{a}}},\tilde{{\mathsf{b}}}}|}},\\
{{\tau}}^{{{\mc{A}},{\mc{B}}}{\tilde{{\mc{A}}},\tilde{{\mc{B}}}}}_{{{{\mathsf{a}}},{{\mathsf{b}}}}{\tilde{{\mathsf{a}}},\tilde{{\mathsf{b}}}}}={{\ms{T}}}_{{\tilde{{\mathsf{a}}},\tilde{{\mathsf{b}}}}{{{\mathsf{a}}},{{\mathsf{b}}}}}\sqrt{\frac{|{\velG}_{{{{\mathsf{a}}},{{\mathsf{b}}}}}|}{|{\velG}_{{\tilde{{\mathsf{a}}},\tilde{{\mathsf{b}}}}}|}}.
\label{scaamp_sq_altinc}
\end{eqn}

The general solution to the equation of motion, at a fixed frequency ${\upomega}$, has the form
\begin{eqn}
{\mathtt{u}}
=\widetilde{{\mathsf{I}}}_{{\mc{I}}}{\mathtt{u}}_{\tilde{{{\mathsf{i}}}}}+\widetilde{{\mathsf{I}}}_{{\mc{A}}}{\mathtt{u}}_{\tilde{{\mathsf{a}}}}+\widetilde{{\mathsf{I}}}_{{\mc{B}}}{\mathtt{u}}_{\tilde{{\mathsf{b}}}}.
\end{eqn}
The asymptotic form of ${\mathtt{u}}
$ is 
\begin{eqn}
\begin{cases}
\widetilde{{\mathsf{I}}}_{{\mc{I}}}{\mathtt{u}}_{{\tilde{{{\mathsf{i}}}}}}+\sum\limits_{{{{\mathsf{i}}}}=1}^{N^{{\mc{I}}}}(\widetilde{{\mathsf{I}}}_{{\mc{I}}}{{\tau}}_{{{\mathsf{i}}}{{\tilde{{{\mathsf{i}}}}}}}\sqrt{\frac{|{\velG}_{{\tilde{{{\mathsf{i}}}}}}|}{|{\velG}_{{{\mathsf{i}}}}|}}+\widetilde{{\mathsf{I}}}_{{\mc{A}}}{{\tau}}_{{{\mathsf{i}}}{\tilde{{\mathsf{a}}}}}\sqrt{\frac{|{\velG}_{\tilde{{\mathsf{a}}}}|}{|{\velG}_{{{\mathsf{i}}}}|}}+\widetilde{{\mathsf{I}}}_{{\mc{B}}}{{\tau}}_{{{\mathsf{i}}}{\tilde{{\mathsf{b}}}}}\sqrt{\frac{|{\velG}_{\tilde{{\mathsf{b}}}}|}{|{\velG}_{{{\mathsf{i}}}}|}}){\mathtt{u}}_{{{\mathsf{i}}}}
\\
\text{ in the terminal ${\mc{I}}$}
\\
\widetilde{{\mathsf{I}}}_{{\mc{A}}}{\mathtt{u}}_{\tilde{{\mathsf{a}}}}+\sum\limits_{{{\mathsf{a}}}=1}^{N^{{\mc{A}}}}(\widetilde{{\mathsf{I}}}_{{\mc{I}}}{{\tau}}_{{{\mathsf{a}}}{\tilde{{{\mathsf{i}}}}}}\sqrt{\frac{|{\velG}_{\tilde{{{\mathsf{i}}}}}|}{|{\velG}_{{\mathsf{a}}}|}}+\widetilde{{\mathsf{I}}}_{{\mc{A}}}{{\tau}}_{{{\mathsf{a}}}{\tilde{{\mathsf{a}}}}}\sqrt{\frac{|{\velG}_{{\tilde{{\mathsf{a}}}}}|}{|{\velG}_{{\mathsf{a}}}|}}+\widetilde{{\mathsf{I}}}_{{\mc{B}}}{{\tau}}_{{{\mathsf{a}}}{\tilde{{\mathsf{b}}}}}\sqrt{\frac{|{\velG}_{{\tilde{{\mathsf{b}}}}}|}{|{\velG}_{{\mathsf{a}}}|}}){\mathtt{u}}_{{\mathsf{a}}}\\
\text{ in the terminal ${\mc{A}}$}
\\
\widetilde{{\mathsf{I}}}_{{\mc{B}}}{\mathtt{u}}_{\tilde{{\mathsf{b}}}}+\sum\limits_{{{\mathsf{b}}}=1}^{N^{{\mc{B}}}}(\widetilde{{\mathsf{I}}}_{{\mc{I}}}{{\tau}}_{{{\mathsf{b}}}{\tilde{{{\mathsf{i}}}}}}\sqrt{\frac{|{\velG}_{\tilde{{{\mathsf{i}}}}}|}{|{\velG}_{{\mathsf{b}}}|}}+\widetilde{{\mathsf{I}}}_{{\mc{A}}}{{\tau}}_{{{\mathsf{b}}}{\tilde{{\mathsf{a}}}}}\sqrt{\frac{|{\velG}_{{\tilde{{\mathsf{a}}}}}|}{|{\velG}_{{\mathsf{b}}}|}}+\widetilde{{\mathsf{I}}}_{{\mc{B}}}{{\tau}}_{{{\mathsf{b}}}{\tilde{{\mathsf{b}}}}}\sqrt{\frac{|{\velG}_{{\tilde{{\mathsf{b}}}}}|}{|{\velG}_{{\mathsf{b}}}|}}){\mathtt{u}}_{{\mathsf{b}}}\\
\text{ in the terminal ${\mc{B}}$.}
\end{cases}
\end{eqn}
Following tradition \cite{Nazarovbook2009}, it is appropriate to define 
\beqan
{{\mathsf{O}}}_{{\mc{I}}}&\equiv& \sqrt{|{\velG}_{{{\mathsf{i}}}}|}{\widetilde{{\mathsf{O}}}}_{{\mc{I}}}
= {{\mathsf{I}}}_{{\mc{I}}}{{\tau}}_{{{\mathsf{i}}}\tilde{{{\mathsf{i}}}}}+{{\mathsf{I}}}_{{\mc{A}}}{{\tau}}_{{{\mathsf{i}}}\tilde{{\mathsf{a}}}}+{{\mathsf{I}}}_{{\mc{B}}}{{\tau}}_{{{\mathsf{i}}}\tilde{{\mathsf{b}}}},
\label{Oaeqn}\\
{{\mathsf{O}}}_{{\mc{A}}}&\equiv& \sqrt{|{\velG}_{{\mathsf{a}}}|}{\widetilde{{\mathsf{O}}}}_{{\mc{A}}}
= {{\mathsf{I}}}_{{\mc{I}}}{{\tau}}_{{{\mathsf{a}}}\tilde{{{\mathsf{i}}}}}+{{\mathsf{I}}}_{{\mc{A}}}{{\tau}}_{{{\mathsf{a}}}\tilde{{\mathsf{a}}}}+{{\mathsf{I}}}_{{\mc{B}}}{{\tau}}_{{{\mathsf{a}}}\tilde{{\mathsf{b}}}},
\label{Oueqn}\\
{{\mathsf{O}}}_{{\mc{B}}}&\equiv& \sqrt{|{\velG}_{{\mathsf{b}}}|}{\widetilde{{\mathsf{O}}}}_{{\mc{B}}}
= {{\mathsf{I}}}_{{\mc{I}}}{{\tau}}_{{{\mathsf{b}}}\tilde{{{\mathsf{i}}}}}+{{\mathsf{I}}}_{{\mc{A}}}{{\tau}}_{{{\mathsf{b}}}\tilde{{\mathsf{a}}}}+{{\mathsf{I}}}_{{\mc{B}}}{{\tau}}_{{{\mathsf{b}}}\tilde{{\mathsf{b}}}},
\label{Oleqn}
\eeqan
where 
$${{\mathsf{O}}}_{{\mc{I}},{{\mc{A}}},{{\mc{B}}}} \equiv \sqrt{|{\velG}_{{{\mathsf{i}}},{{\mathsf{a}}},{{\mathsf{b}}}}|}{\widetilde{{\mathsf{O}}}}_{{\mc{I}},{{\mc{A}}},{{\mc{B}}}},
$$ and $$
{{\mathsf{I}}}_{{\mc{I}},{{\mc{A}}}, {{\mc{B}}}}\equiv \sqrt{|{\velG}_{\tilde{{{\mathsf{i}}}},\tilde{{\mathsf{a}}},\tilde{{\mathsf{b}}}}|}{\widetilde{{\mathsf{I}}}}_{{\mc{I}},{{\mc{A}}}, {{\mc{B}}}}$$ {can be easily identified} as the flux amplitudes.
Note that ${\mathtt{u}}_{{{\mathsf{i}}}}$ represents a wave moving outward from the nanojunction, while ${\mathtt{u}}_{\tilde{{\mathsf{i}}}}$ is a wave moving inward towards the nanojunction, deep in ${{\mc{A}}}$ lead, analog holds for both ${\mc{I}}$ and ${{\mc{B}}}$ leads. Thus, above equations \eqref{Oaeqn}--\eqref{Oleqn} define a linear relation between the flux amplitudes of outgoing and incoming waves. 
This relation can be written in a matrix form using the matrix $\Ttt{S}$, that relates the outgoing flux amplitudes ${{\mathsf{O}}}_{{\mc{I}},{\mc{A}},{\mc{B}}}$ to the incoming flux amplitudes ${{\mathsf{I}}}_{{\mc{I}},{\mc{A}},{\mc{B}}}$ 
 \cite{Nazarovbook2009}. 
The $\Ttt{S}$ matrix is a $$(N^{{\mc{I}}}+N^{{\mc{A}}}+N^{{\mc{B}}}) \times (N^{{\mc{I}}}+N^{{\mc{A}}}+N^{{\mc{B}}})$$ matrix. 
Using the labels of these channels denoted by ${{\mathsf{i}}}=1, 2, \dotsc, N^{{\mc{I}}}, {{{\mathsf{a}}},{{\mathsf{b}}}} = 1, 2, \dotsc, N^{{{\mc{A}},{\mc{B}}}}$, 
with
$$\Tt{{\mathsf{O}}}=[{{\mathsf{O}}}^1_{{\mc{I}}}~\cdots~{{\mathsf{O}}}^{N^{{\mc{I}}}}_{{\mc{I}}}~{{\mathsf{O}}}^1_{{\mc{A}}}~
\cdots~{{\mathsf{O}}}^{N^{{\mc{A}}}}_{{\mc{A}}}~
{{\mathsf{O}}}^1_{{\mc{B}}}~\cdots~{{\mathsf{O}}}^{N^{{\mc{B}}}}_{{\mc{B}}}]^T$$ and $$\Tt{{\mathsf{I}}}=[{{\mathsf{I}}}^1_{{\mc{I}}}~\cdots~{{\mathsf{I}}}^{N^{{\mc{I}}}}_{{\mc{I}}}~{{\mathsf{I}}}^1_{{\mc{A}}}~
\cdots~{{\mathsf{I}}}^{N^{{\mc{A}}}}_{{\mc{A}}}~
{{\mathsf{I}}}^1_{{\mc{B}}}~\cdots~{{\mathsf{I}}}^{N^{{\mc{B}}}}_{{\mc{B}}}]^T,$$
above has the general form
$\Tt{{\mathsf{O}}}=\Ttt{S}\Tt{{\mathsf{I}}}.$
In particular, the coefficient matrix can be expressed as
\begin{eqn}\Ttt{S}=\begin{bmatrix}
{{\tau}}^{{\mc{I}}\tilde{{\mc{I}}}}_{N^{{\mc{I}}}\times N^{{\mc{I}}}}&{{\tau}}^{{\mc{I}}{\tilde{{\mc{A}}}}}_{N^{{\mc{I}}}\times N^{{\mc{A}}}}&{{\tau}}^{{\mc{I}}{\tilde{{\mc{B}}}}}_{N^{{\mc{I}}}\times N^{{\mc{B}}}}\\
{{\tau}}^{{{\mc{A}}}\tilde{{\mc{I}}}}_{N^{{\mc{A}}}\times N^{{\mc{I}}}}&{{\tau}}^{{{\mc{A}}}{\tilde{{\mc{A}}}}}_{N^{{\mc{A}}}\times N^{{\mc{A}}}}&{{\tau}}^{{{\mc{A}}}{\tilde{{\mc{B}}}}}_{N^{{\mc{A}}}\times N^{{\mc{B}}}}\\
{{\tau}}^{{{\mc{B}}}\tilde{{\mc{I}}}}_{N^{{\mc{B}}}\times N^{{\mc{I}}}}&{{\tau}}^{{{\mc{B}}}{\tilde{{\mc{A}}}}}_{N^{{\mc{B}}}\times N^{{\mc{A}}}}&{{\tau}}^{{{\mc{B}}}{\tilde{{\mc{B}}}}}_{N^{{\mc{B}}}\times N^{{\mc{B}}}}
\end{bmatrix}.
\end{eqn}
Recalling the relations between the transmission and reflection probabilities, it is easy to show that the $\Ttt{S}$ matrix is unitary, i.e., 
$\Ttt{S}\Ttt{S}^{\dagger}=\Ttt{S}^{\dagger}\Ttt{S}=\Ttt{1},$
which is another way of expressing particle flux conservation \cite{Nazarovbook2009}. 
As a necessary consequence of the unitarity it follows that
\begin{eqn}
\sum\nolimits_{{{{\mathsf{i}}}}=1}^{N^{{\mc{I}}}}|{{\tau}}^{{\mc{I}}\tilde{{\mc{I}}}}_{{{\mathsf{i}}}\tilde{{{\mathsf{i}}}}}|^2+\sum\nolimits_{{{\mathsf{a}}}=1}^{N^{{\mc{A}}}}|{{\tau}}^{{\mc{A}}\tilde{{\mc{I}}}}_{{\mathsf{a}}\tilde{{{\mathsf{i}}}}}|^2+\sum\nolimits_{{{\mathsf{b}}}=1}^{N^{{\mc{B}}}}|{{\tau}}^{{\mc{B}}\tilde{{\mc{I}}}}_{{\mathsf{b}}\tilde{{{\mathsf{i}}}}}|^2=1, \\
\sum\nolimits_{{{{\mathsf{i}}}}=1}^{N^{{\mc{I}}}}|{{\tau}}^{{\mc{I}}\tilde{{\mc{A}}}}_{{{\mathsf{i}}}\tilde{{\mathsf{a}}}}|^2+\sum\nolimits_{{{\mathsf{a}}}=1}^{N^{{\mc{A}}}}|{{\tau}}^{{\mc{A}}\tilde{{\mc{A}}}}_{{\mathsf{a}}\tilde{{\mathsf{a}}}}|^2+\sum\nolimits_{{{\mathsf{b}}}=1}^{N^{{\mc{B}}}}|{{\tau}}^{{\mc{B}}\tilde{{\mc{A}}}}_{{{\mathsf{b}}}\tilde{{\mathsf{a}}}}|^2=1,\\
\sum\nolimits_{{{{\mathsf{i}}}}=1}^{N^{{\mc{I}}}}|{{\tau}}^{{\mc{I}}\tilde{{\mc{B}}}}_{{{\mathsf{i}}}\tilde{{\mathsf{b}}}}|^2+\sum\nolimits_{{{\mathsf{a}}}=1}^{N^{{\mc{A}}}}|{{\tau}}^{{\mc{A}}\tilde{{\mc{B}}}}_{{\mathsf{a}}\tilde{{\mathsf{b}}}}|^2+\sum\nolimits_{{{\mathsf{b}}}=1}^{N^{{\mc{B}}}}|{{\tau}}^{{\mc{B}}\tilde{{\mc{B}}}}_{{\mathsf{b}}\tilde{{\mathsf{b}}}}|^2=1,
\label{energybalanceanalog}
\end{eqn}
for arbitrary $\tilde{{{\mathsf{i}}}}$, $\tilde{{\mathsf{a}}}$, and $\tilde{{\mathsf{b}}}$, respectively, corresponding to the incidence from the respective portions ${\mc{I}},{\mc{A}},$ and ${\mc{B}}$. 
\eqref{energybalanceanalog} can be identified as the statement of energy balance for the mechanical lattice model (without dissipation). The equation \eqref{energybalanceanalog}${}_1$ has been established as the zero lemma by 
\paperone{} for the incidence from the direction of portion ${\mc{I}}$ (the other two cases of incidence from ${\mc{A}}$ and ${\mc{B}}$ are analogous),
i.e. ${\ms{R}}={\ms{R}}_{{\mc{I}}}$ and ${\ms{T}}={\ms{T}}_{{\mc{A}}}+{\ms{T}}_{{\mc{B}}}$ defined by \eqref{Ref_sq_k} and \eqref{Trans_sq_kup}\&\eqref{Trans_sq_klow} (and \eqref{Ref_sq_c} and \eqref{Trans_sq_cup}\&\eqref{Trans_sq_clow}) for the 
ribbons, satisfy the condition 
${\ms{R}}+{\ms{T}}=1.$
In view of \eqref{energyflux_inc}, the statement of the zero lemma stated in \paperone 
can be re-written as
${\ms{E}^{i}}(\tilde{{\mathsf{i}}}){\ms{R}}_{{\mc{I}}}(\tilde{{\mathsf{i}}})+{\ms{E}^{i}}(\tilde{{\mathsf{i}}}){\ms{T}}_{{\mc{A}}}(\tilde{{\mathsf{i}}})+{\ms{E}^{i}}(\tilde{{\mathsf{i}}}){\ms{T}}_{{\mc{B}}}(\tilde{{\mathsf{i}}})={\ms{E}^{i}}(\tilde{{\mathsf{i}}}),$
${\ms{E}^{i}}(\tilde{{\mathsf{a}}}){\ms{T}}_{{\mc{I}}}(\tilde{{\mathsf{a}}})+{\ms{E}^{i}}(\tilde{{\mathsf{a}}}){\ms{R}}_{{\mc{A}}}(\tilde{{\mathsf{a}}})+{\ms{E}^{i}}(\tilde{{\mathsf{a}}}){\ms{T}}_{{\mc{B}}}(\tilde{{\mathsf{a}}})={\ms{E}^{i}}(\tilde{{\mathsf{a}}}),$
${\ms{E}^{i}}(\tilde{{\mathsf{b}}}){\ms{T}}_{{\mc{I}}}(\tilde{{\mathsf{b}}})+{\ms{E}^{i}}(\tilde{{\mathsf{b}}}){\ms{T}}_{{\mc{A}}}(\tilde{{\mathsf{b}}})+{\ms{E}^{i}}(\tilde{{\mathsf{b}}}){\ms{R}}_{{\mc{B}}}(\tilde{{\mathsf{b}}})={\ms{E}^{i}}(\tilde{{\mathsf{b}}}),$
for arbitrary $\tilde{{{\mathsf{i}}}}$, $\tilde{{\mathsf{a}}}$, and $\tilde{{\mathsf{b}}}$, respectively, corresponding to the incidence from the respective portions ${\mc{I}},{\mc{A}},$ and ${\mc{B}}$. 

\section{Chebyshev polynomials}
\label{app_Chebpoly}
Using the zeros of Chebyshev polynomials \cite{MasonHand}, in particular, Table 3 and Eq. (100) of a recent paper \cite{Bls9}, the dispersion relations for the elementary three kinds of outer boundary conditions for square lattice waveguides can be written in terms of
\begin{eqn}
{\mathtt{U}}_n&=\prod\nolimits_{j=1}^{n}({{\mathpzc{H}}}+4\sin^2\tfrac{j}{2(n+1)}\pi),\\
{\mathtt{V}}_n&=\prod\nolimits_{j=1}^{n}({{\mathpzc{H}}}+4\sin^2\tfrac{j-{\frac{1}{2}}}{2n+1}\pi),\\
{\mathtt{W}}_n&=\prod\nolimits_{j=1}^{n}({{\mathpzc{H}}}+4\sin^2\tfrac{j}{2n+1}\pi),\\
{\mathtt{T}}_n&=\prod\nolimits_{j=1}^{n}({{\mathpzc{H}}}+4\sin^2\tfrac{j-{\frac{1}{2}}}{2n}\pi),
\label{chebpoly}
\end{eqn}
where ${\mathpzc{H}}$ is defined in \eqref{dHelmholtzF_sq} (also recall 
that ${\mathpzc{H}}=2({\vartheta}-1).$).
Using Table \ref{bifurcatedstripbc_sq_ND} and \eqref{chebpoly} 
the `transparent' wave modes can be identified by seeking the common factors in the three possible pairs of the sets as listed in each of the six cases (recall Fig. \ref{latticestrip_sq_BCs_asymm_semiinfdefect}) of bifurcation:
\begin{description}
\item[(a)] 
 $\{\frac{j}{{\mathtt{N}}+1}\}_{j=1}^{{\mathtt{N}}},$ $\{\frac{j-{\frac{1}{2}}}{{\mathtt{N}_{\mathfrak{a}}}+{\frac{1}{2}}}\}_{j=1}^{{\mathtt{N}_{\mathfrak{a}}}},$ $ \{\frac{j-{\frac{1}{2}}}{{\mathtt{N}_{\mathfrak{b}}}+{\frac{1}{2}}}\}_{j=1}^{{\mathtt{N}_{\mathfrak{b}}}}$;
\item[(b)] $\{\frac{j}{{\mathtt{N}}}\}_{j=0}^{{\mathtt{N}}-1},$ $ \{\frac{j}{{\mathtt{N}_{\mathfrak{a}}}}\}_{j=0}^{{\mathtt{N}_{\mathfrak{a}}}-1},$ $ \{\frac{j}{{\mathtt{N}_{\mathfrak{b}}}}\}_{j=0}^{{\mathtt{N}_{\mathfrak{b}}}-1}$;
\item[(c)] $\{\frac{j-{\frac{1}{2}}}{{\mathtt{N}}+{\frac{1}{2}}}\}_{j=1}^{{\mathtt{N}}},$ $ \{\frac{j-{\frac{1}{2}}}{{\mathtt{N}_{\mathfrak{a}}}+{\frac{1}{2}}}\}_{j=1}^{{\mathtt{N}_{\mathfrak{a}}}},$ $ \{\frac{j}{{\mathtt{N}_{\mathfrak{b}}}}\}_{j=0}^{{\mathtt{N}_{\mathfrak{b}}}-1}$;
\end{description}
and
\begin{description}
\item[(a')] $\{\frac{j}{{\mathtt{N}}+1}\}_{j=1}^{{\mathtt{N}}},$ $ \{\frac{j}{{\mathtt{N}_{\mathfrak{a}}}+1}\}_{j=1}^{{\mathtt{N}_{\mathfrak{a}}}},$ $ \{\frac{j}{{\mathtt{N}_{\mathfrak{b}}}+1}\}_{j=1}^{{\mathtt{N}_{\mathfrak{b}}}}$;
\item[(b')] $\{\frac{j}{{\mathtt{N}}}\}_{j=0}^{{\mathtt{N}}-1},$ $ \{\frac{j-{\frac{1}{2}}}{{\mathtt{N}_{\mathfrak{a}}}+{\frac{1}{2}}}\}_{j=1}^{{\mathtt{N}_{\mathfrak{a}}}},$ $ \{\frac{j-{\frac{1}{2}}}{{\mathtt{N}_{\mathfrak{b}}}+{\frac{1}{2}}}\}_{j=1}^{{\mathtt{N}_{\mathfrak{b}}}}$;
\item[(c')] $\{\frac{j-{\frac{1}{2}}}{{\mathtt{N}}+{\frac{1}{2}}}\}_{j=1}^{{\mathtt{N}}},$ $ \{\frac{j}{{\mathtt{N}_{\mathfrak{a}}}+1}\}_{j=1}^{{\mathtt{N}_{\mathfrak{a}}}},$ $ \{\frac{j-{\frac{1}{2}}}{{\mathtt{N}_{\mathfrak{b}}}+{\frac{1}{2}}}\}_{j=1}^{{\mathtt{N}_{\mathfrak{b}}}}$.
\end{description}

\end{appendix}

\section*{References}
\bibliography{}

\end{document}